\documentclass[%
reprint,
superscriptaddress,
 amsmath,amssymb,
 aps,
]{revtex4-1}

\usepackage{mathtools}
\usepackage{graphicx}
\usepackage{dcolumn}
\usepackage{bm}

\usepackage{amsmath}
\usepackage{amssymb}
\usepackage{amsthm}
\usepackage{graphicx}
\usepackage{algorithmic}
\usepackage{mathrsfs}
\usepackage{braket}
\usepackage{xcolor}
\usepackage{tikz}
\usepackage[roman]{complexity}
\usepackage[caption=false]{subfig}

\usepackage{pdfpages}
\usepackage{pgffor}
\makeatletter
\AtBeginDocument{\let\LS@rot\@undefined}
\makeatother

\begin{document}

\title{Estimating the randomness of quantum circuit ensembles up to 50 qubits}

\author{Minzhao Liu}
\email{mliu6@uchicago.edu}
\affiliation{Department of Physics, The University of Chicago, Chicago, IL 60637, USA}
\affiliation{Computational Science Division, Argonne National Laboratory, Lemont, IL 60439, USA}

\author{Junyu Liu$^*$}
\email{junyuliu@uchicago.edu}
\affiliation{Pritzker School of Molecular Engineering, The University of Chicago, Chicago, IL 60637, USA}
\affiliation{Chicago Quantum Exchange, Chicago, IL 60637, USA}
\affiliation{Kadanoff Center for Theoretical Physics, The University of Chicago, Chicago, IL 60637, USA}
\affiliation{qBraid Co., Harper Court 5235, Chicago, IL 60615, USA}

\author{Yuri Alexeev}
\email{yuri@anl.gov}
\affiliation{Computational Science Division, Argonne National Laboratory, Lemont, IL 60439, USA}
\affiliation{Chicago Quantum Exchange, Chicago, IL 60637, USA}

\author{Liang Jiang}
\email{liangjiang@uchicago.edu}
\affiliation{Pritzker School of Molecular Engineering, The University of Chicago, Chicago, IL 60637, USA}
\affiliation{Chicago Quantum Exchange, Chicago, IL 60637, USA}

\date{\today}

\begin{abstract}
Random quantum circuits have been utilized in the contexts of quantum supremacy demonstrations, variational quantum algorithms for chemistry and machine learning, and blackhole information. The ability of random circuits to approximate any random unitaries has consequences on their complexity, expressibility, and trainability. To study this property of random circuits, we develop numerical protocols for estimating the frame potential, the distance between a given ensemble and the exact randomness. Our tensor-network-based algorithm has polynomial complexity for shallow circuits and is high-performing using CPU and GPU parallelism. We study 1. local and parallel random circuits to verify the linear growth in complexity as stated by the Brown--Susskind conjecture, and; 2. the hardware-efficient ans{\"a}tze to shed light on its expressibility and the barren plateau problem in the context of variational algorithms. Our work shows that large-scale tensor network simulations could provide important hints toward open problems in quantum information science.

\end{abstract}

\maketitle


$*$: Corresponding author.

\section{Introduction}
Quantum computing might provide significant improvement of computational powers for current information technologies \cite{feynman1982simulating,preskill2012quantum,alexeev2021quantum}. In the noisy intermediate-scale quantum (NISQ) era, an important question for near-term quantum computing is whether quantum devices are able to realize strong computational advantage against existing classical devices and resolve hard problems that no existing classical computers can resolve \cite{preskill2018quantum}. Recently, Google and the University of Science and Technology of China, in  experiments involving boson sampling \cite{arute2019quantum,zhong2020quantum}, claimed to have  realized  quantum advantage using their quantum devices, disproving the extended Church--Turing thesis.  These experiments are considered milestones toward full-scale quantum computing. Another recent study suggests the possibility of achieving  quantum advantage in runtime over specialized state-of-the-art heuristic algorithms to solve the Maximum Independent Set problem using Rydberg atom arrays \cite{ebadi2022quantum}.

 Despite the great experimental success in quantum devices, however, the capability of classical computation is also rapidly developing. It is interesting and important to think about where the boundary of classical computation of the same process is and to understand the underlying physics of the quantum supremacy experiments through classical simulation \cite{arute2019quantum}. Tensor network methods are incredibly useful for simulating quantum circuits \cite{white1992density,rommer1997class,orus2019tensor}. Originating from approximately solving ground states of quantum many-body systems, tensor network methods find approximate solutions when the bond dimension of contracted tensors and the required entanglement of the system is under control \cite{white1992density}. Tensor network methods are also widely used for investigating sampling experiments with random quantum architectures, which are helpful for verifying the quantum supremacy experiments \cite{noh2020efficient,huang2020classical,pan2021simulating,oh2021classical}. 

In this work, we develop novel tensor network methods and perform classical random circuit sampling experiments up to 50 qubits. Random circuit sampling experiments are important components of near-term characterizations of quantum advantage \cite{boixo2018characterizing}. Ensembles of random circuits could provide implementable constructions of approximate unitary $k$-designs \cite{harrow2009random,brandao2016local,stat}, quantum information scramblers \cite{Hayden:2007cs}, solvable many-body physics models \cite{Nahum:2017yvy}, predictable variational ans\"{a}tze for quantum machine learning \cite{mcclean2018barren,Liu:2021wqr,Liu:2022eqa}, good quantum decouplers for quantum channel and quantum error correction codes \cite{brown2015decoupling,Liu:2020sqb}, and efficient representatives of quantum randomness. To measure how close a given random circuit ensemble is to Haar-uniform randomness over the unitary group, we develop algorithms to evaluate the frame potential, the 2-norm distance toward full Haar randomness \cite{Roberts:2016hpo,Cotler:2017jue,Liu:2018hlr}. The frame potential is a user-friendly measure of how random a given ensemble is in terms of operator norms: the smaller the frame potential is, the more chaotic and more complicated the ensembles are, and the more easily we can achieve computational advantages \cite{brandao2010exponential,harlow2013quantum}. In fact, in certain quantum cryptographic tools,  concepts identical or similar to approximate $k$-designs are used,  making use of the exponential separation of complexities between classical and quantum computations \cite{brandao2016efficient,ji2017pseudorandom,ananth2021cryptography,vskoric2012quantum,gianfelici2020theoretical,kumar2021efficient,doosti2021connection,arapinis2021quantum}.

\textcolor{black}{It is critical to perform simulations of quantum circuits efficiently. To achieved this, we developed an efficient tensor network contraction algorithm is developed in the \texttt{QTensor} package \cite{lykov2021performance,lykov_diagonal,lykov2021large}. QTensor package is optimized to simulate large quantum circuits on supercomputers. For this project, we implemented a modified tensor network and fully utilized QTensor's ability to simulate quantum circuits efficiently at scale.}

In particular, we show the following applications of our computational tools. First, we evaluate the $k$-design time of the local and parallel random circuits through the frame potential. A long-term open problem is to prove the linear scrambling property of random circuits, where they approach approximate $k$-designs at depth $\mathcal{O} (nk)$ with $n$ qudits \cite{harrow2009random,brandao2010exponential,diniz2011comment,brandao2016local,brandao2016efficient,harrow2018approximate,nakata2016efficient,onorati2017mixing,Lashkari:2011yi,stat,Brandao:2019sgy}. Although lower and upper bounds are given, there is no known proof of the $k$-design time for general local dimension $q$ and $k \ge 3$ \cite{stat,Brandao:2019sgy}. According to \cite{Brandao:2019sgy}, the linear increase of the $k$-design time will lead to a proof of the Brown--Susskind conjecture, a statement where random circuits have linear growth of the circuit complexity with insights from black hole physics \cite{Brown:2017jil,Susskind:2018fmx}. Recently, the complexity statement was proved in \cite{haferkamp2022linear} for a different definition of circuit complexity compared with   \cite{Brandao:2019sgy}. Thus, a validation of the $k$-design time measured in the frame potential will immediately lead to an alternative verification of the Brown--Susskind conjecture, with the complexity defined in \cite{Brandao:2019sgy}. Using our tools, we verify the linear scaling of the $k$-design time up to 50 qubits and $q=2$. Our research also provides important data on the prefactors beyond scaling through numerical simulations, which will be helpful to further the understanding of theoretical computer scientists. 

Moreover, we use our tools to evaluate the frame potential of the randomized hardware-efficient variational ans{\"a}tze used in \cite{mcclean2018barren}. \emph{Barren plateau} is a term referring to the slowness of the variational angle updates during the gradient descent dynamics of quantum machine learning. When the variational ans\"{a}tze for variational quantum simulation, variational quantum optimization, and quantum machine learning \cite{peruzzo2014variational,mcclean2016theory,kandala2017hardware,cerezo2021variational,farhi2014quantum,wittek2014quantum,wiebe2014quantum,biamonte2017quantum,schuld2019quantum,havlivcek2019supervised,liu2021rigorous,Liu:2021ohs,farhi2018classification} are random enough, the gradient descent updates of variational angles will be suppressed by the dimension of Hilbert space, requiring exponential precision to implement quantum control of variational angles \cite{Liu:2022eqa}. The quadratic fluctuations considered in \cite{mcclean2018barren,mcclean2018barren} will be suppressed with an assumption of 2-design, which is claimed to be satisfied by their hardware-efficient variational ans\"{a}tze. For higher moments, higher $k$-designs are required. A study of how far from a given variational assumption to a unitary $k$-design is important in order to understand how large the barren plateau is and how to mitigate them through designs of variational circuits. In our work, we find that for several $k$s, the randomized hardware-efficient ans\"{a}tze are efficient scramblers: the frame potential decays exponentially during an increase in circuit depth, \textcolor{black}{and non-diagonal entangling gates are more efficient.}

\subsection{Frame potential}\label{frame}

Given an ensemble $\mathcal{E}$ of unitaries with a probability measure, we are interested in its randomness and, therefore, closeness to the unitary group. Truly random unitaries from the unitary group have the \textit{Haar} measure. Such closeness is measured by how well the ensemble approximates the first $k$ moments of the unitary group. To this end, a $k$-fold twirling channel

\begin{equation}
    \Phi_{\mathcal{E}}^{(k)}(\mathcal{O})=\int_{\mathcal{E}}dUU^{\otimes k}(\mathcal{O})U^{\dagger\otimes k}
\end{equation}
is defined for the ensemble. If the unitary ensemble approximates the $k$th moment of the unitary group, the distance between the $k$-fold channel defined for the ensemble and the Haar unitaries (measured by the diamond norm) is bounded by $\epsilon$:
\begin{equation}
    \|\Phi_{\mathcal{E}}^{(k)}-\Phi_{\text{Haar}}^{(k)}\|_{\diamond}\leq\epsilon.
\end{equation}

Such $\mathcal{E}$ is said to be an $\epsilon$-approximate $k$-design. The diamond norm of the channels is not numerically friendly, however. A quantity more suitable for numerical evaluation, which is also discussed in the context of $k$-designs, is the \textit{frame potential} $\mathcal{F}$, given by \cite{frame_potential}
\begin{equation}
    \mathcal{F}_{\mathcal{E}}^{(k)}=\int_{U,V\in\mathcal{E}}dUdV\vert \text{Tr}(U^{\dagger}V)\vert^{2k}.
\end{equation}

Specifically, it relates to the aforementioned definition of $\epsilon$-approximate $k$-designs as follows \cite{stat}:
\begin{equation}
    \|\Phi_{\mathcal{E}}^{(k)}-\Phi_{\text{Haar}}^{(k)}\|_{\diamond}^2\leq d^{2k}(\mathcal{F}_{\mathcal{E}}^{(k)}-\mathcal{F}_{\text{Haar}}^{(k)}),
\end{equation}
where $d=q^n$ is the Hilbert space dimension, $q$ is the local dimension of the qudits, and $\mathcal{F}_{\text{Haar}}^{(k)}=k!$.

If we obtain the frame potential $\mathcal{F}_{\mathcal{E}}^{(k)}$, we are guaranteed to have  at least an $\epsilon_{\text{max}}$-approximate $k$-design, where
\begin{equation}
    \epsilon_{\text{max}}=d^k\sqrt{\mathcal{F}_{\mathcal{E}}^{(k)}-\mathcal{F}_{\text{Haar}}^{(k)}}\label{epsilonmax},
\end{equation}

Similarly, we have the following condition for the ensemble to be an $\epsilon$-approximate $k$-design:

\begin{equation}
    \sqrt{\mathcal{F}_{\mathcal{E}(l)}^{(k)}-\mathcal{F}_{\text{Haar}}^{(k)}}\leq \frac{\epsilon}{q^{nk}},
\end{equation}
where the ensemble $\mathcal{E}(l)$ depends on the number of layers $l$. Assuming an exponentially decreasing frame potential approaching the Haar value, we have
\begin{align}
    & \mathcal{F}_{\mathcal{E}(l)}^{(k)}-\mathcal{F}_{\text{Haar}}^{(k)} \propto A^2 e^{-2l/C}\label{exponential}\\
    \Rightarrow & A e^{-l/C}\leq \frac{\epsilon}{q^{nk}}\\
    \Rightarrow & l\geq C(kn\log{q}+\log{A}+\log{1/\epsilon})\label{layers}.
\end{align}

Under this assumption, $A$ and $C$ could still have $n$ and $k$ dependence. Therefore, for there to be linear scaling in $n$ and $k$, $A$ cannot be exponential, and $C$ must be sublinear. 

\begin{figure} [ht]
   \begin{center}
   \includegraphics[width=7cm]{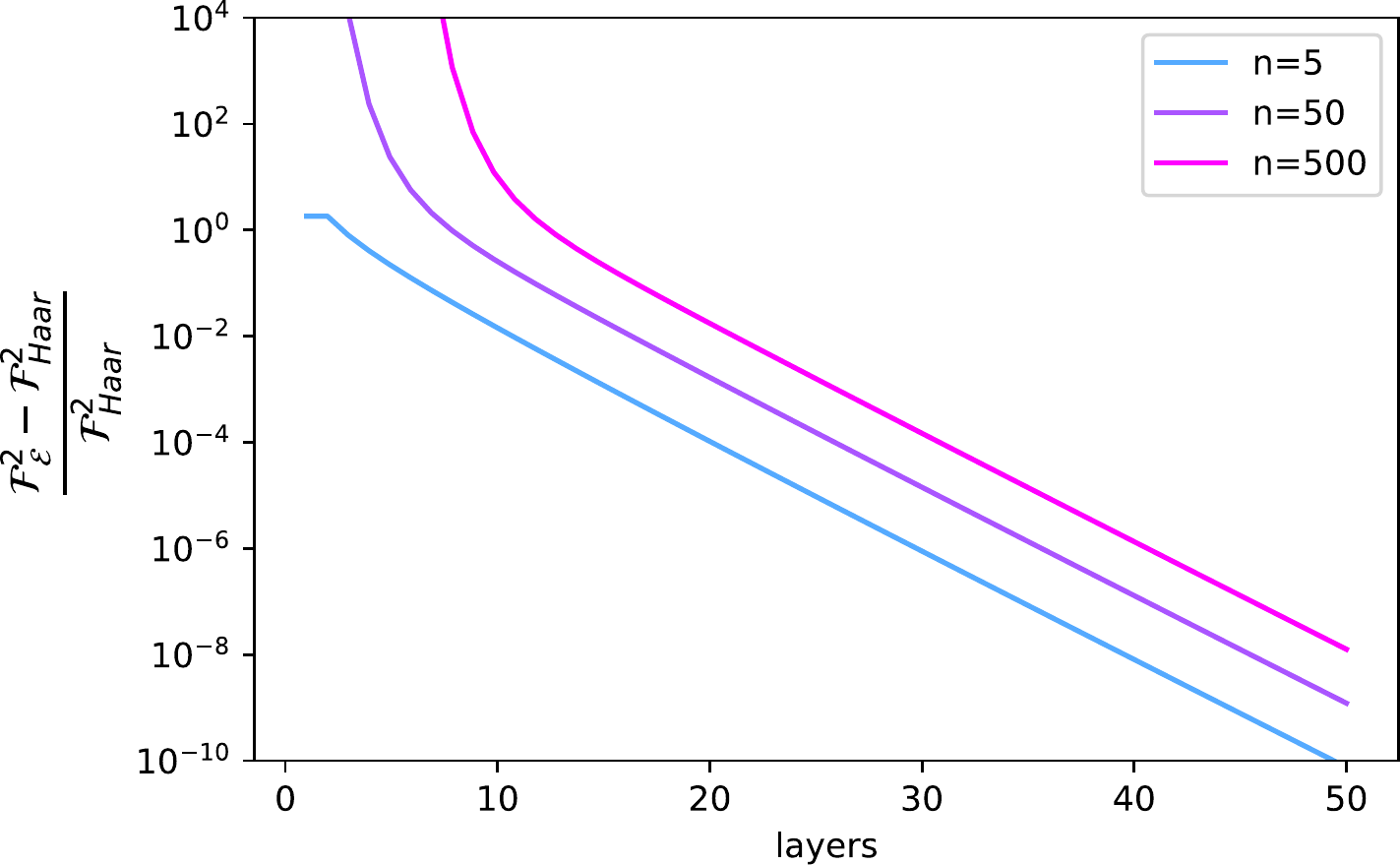}
   \end{center}
   \caption{Theoretical \textcolor{black}{fractional} deviation of the $k=2$ frame potential from the Haar value as a function of layers for the parallel random unitary ans{\"a}tze. In this plot, the layer required to reach a fixed $\mathcal{F}$ does not scale linearly with $n$. 
   The linear scaling is only for fixed $\epsilon$.}
   { \label{F theory}
}
   \end{figure}

As an example, the exponential decay of $\mathcal{F}^{(2)}$ for the parallel random unitary ans{\"a}tze is given by \cite{stat}
\begin{equation}
        \mathcal{F}^{(2)}< 2\left(1+\left(\frac{2q}{q^2+1}\right)^{2(l-1)}\right)^{n_g-1},\label{theory_eq}
\end{equation}
where $n_g=\lfloor n/2\rfloor$. This is plotted in Fig. \ref{F theory}. For fixed $\epsilon$, this leads to a linear scaling of $l$ in $n$, given by
\begin{equation}
    l \geq C(2n\log{q}+\log{n}+\log{1/\epsilon})\label{k=2},
\end{equation}
where $C=\left(\log{\frac{q^2+1}{2q}}\right)^{-1}$ is independent of $n$. We emphasize that linear scaling in $n$ is for fixed $\epsilon$, not fixed $\mathcal{F}$.

\section{Results}\label{result}

\begin{figure*} [ht]
   \begin{center}
   \includegraphics[width=15cm]{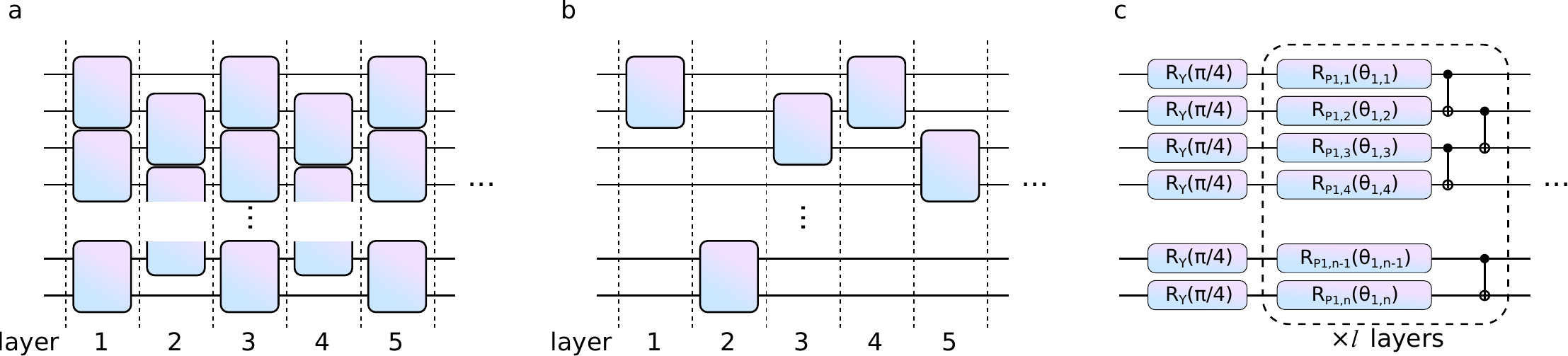}
   \end{center}
   \caption{\textcolor{black}{Illustration of ans{\"a}tzes used in this work. All ans{\"a}tzes assumes 1D nearest-neighbor connectivity.} (a) Parallel random unitary ans{\"a}tze. Each layer is a wall of two-qudit random unitaries on neighboring qudits, and the next layer is offset by 1 qudit. This creates a brickwork motif, and the gate count scales as $O(ln)$. (b) Local random unitary ans{\"a}tze. Each layer is a single two-qudit random unitary between a pair of randomly chosen neighboring qudits. The gate count scales as $O(l)$. (c) Hardware-efficient ans{\"a}tze. A wall of $R_Y(\pi/4)$ rotations is followed by alternating layers of random Pauli rotations and \textcolor{black}{controlled-NOT} gates, all independently parameterized.}
   { \label{random ansatze}
}
   \end{figure*}

We obtain numerical results for ans{\"a}tze with local dimension $q=2$. Specifically, the frame potential values up to 50 qubits and $k=5$ are evaluated. We compute the frame potentials for local random unitary ans{\"a}tze, parallel random unitary ans{\"a}tze and hardware efficient ans{\"a}tze, which are illustrated in Fig. \ref{random ansatze}.

\subsection{Algorithm description}\label{algorithm}

\begin{figure} [ht]
   \begin{center}
   \includegraphics[width=7cm]{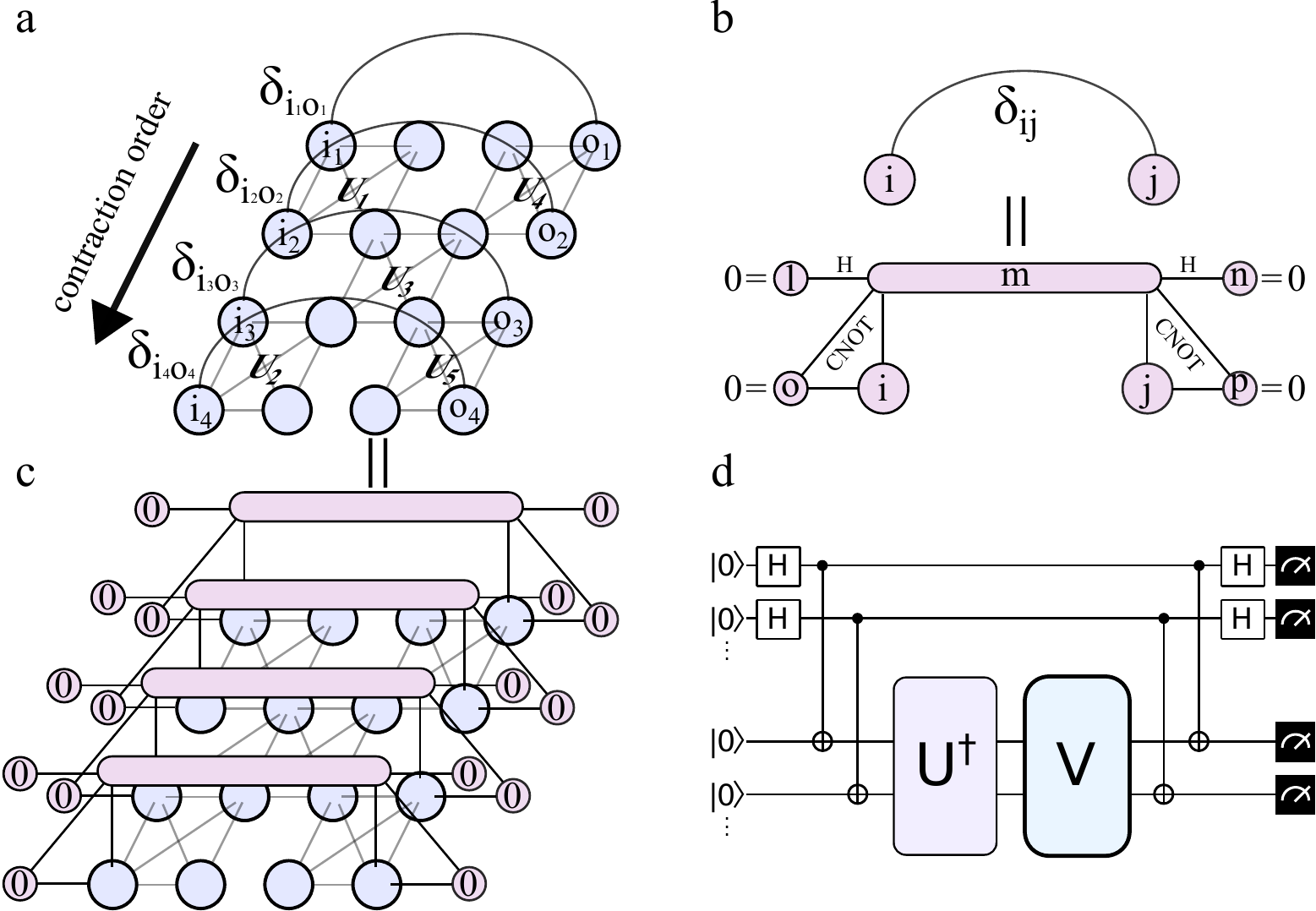}
   \end{center}
   \caption{(a) Graphical tensor network representation of the trace of a quantum circuit, \textcolor{black}{where nodes are tensor indices, and cliques are tensors.} The $i$ indices correspond to qubit inputs, and the $o$ indices correspond to qubit outputs. The curves going above the circuit network are identities. The input and output indices can actually be merged together, but this is harder to illustrate. \textcolor{black}{(b) Equivalent identity tensor that can be represented by gates, qubit initialization, and measurement. More details at the end of section \ref{algorithm}}. (c) Graphical tensor network representation of the same quantity using our formulation. (d) The quantum circuit used to evaluate traces as a single amplitude.}
   { \label{trace}
}
   \end{figure}

The unitary ensembles we are interested in are parameterized by a large number of parameters. Therefore, evaluating the integral is a high-dimensional integration problem, and a numerical Monte Carlo approach is suitable. We approximate the frame potential as the mean value of the trace,
\begin{equation}
    \mathcal{F}_{\mathcal{E}}^{(k)}\approx\frac{1}{N}\sum\vert\text{Tr}(U^{\dagger}V)\vert^{2k}, \ U,V\in\mathcal{E}.
\end{equation}

Therefore, we need to evaluate the trace of the sampled unitaries on $n$ target qudits.

A quantum circuit unitary $U=U_1 U_2 U_3\dots$ is a tensor $U_{ijk\dots}^{\alpha\beta\gamma\dots}$, where $i,j,k$ are input qubit indices and $\alpha,\beta,\gamma$ are output qubit indices. The trace of the unitary is 
\begin{equation}
    \text{Tr}(U)=\sum_{ijk\dots\alpha\beta\gamma\dots}U_{ijk\dots}^{\alpha\beta\gamma\dots}\delta_{i\alpha}\delta_{j\beta}\delta_{k\gamma}\cdots.
\end{equation}
This is a tensor contraction operation that can be expressed as the tensor network in Fig. \ref{trace}a. In this representation, each node is an index, and edges that form cliques are unitaries. \textcolor{black}{This is different from tensor network representations that are more familiar in other works (MPS, MPO, MERA, etc.) For more details on the representation, see \cite{boixo2018characterizing,Schutski,lykov_diagonal}.} The circuit shown here is a parallel random unitary circuit with 4 qubits. For efficient contraction, when the number of qubits is large, the contraction order is along the direction indicated in the Fig. \ref{trace} such that the maximum number of exposed indices is minimum.

Directly implementing this tensor network requires modification of QTensor. We propose an alternative tensor network in Fig. \ref{trace}c with similar topologies that gives the trace as a single-probability amplitude in the form of $\langle\psi\vert U\vert\psi\rangle$ for any basis state $\vert\psi\rangle$. The quantum circuit to achieve this is illustrated in Fig. \ref{trace} c, and we proceed with a proof.

For simplicity, we describe the algorithm for $q=2$ qubits. We assign an ancillary qubit to each target qubit. The quantum state of the $n$ ancillary and $n$ target qubits is initialized to the state
\begin{equation}
    \vert\Psi\rangle_0 = \vert00\cdots0\rangle_a\otimes\vert00\cdots0\rangle_t.
\end{equation}
After a layer of Hadamard on the ancillary qubits, we  get
\begin{align}
    \vert\Psi\rangle&\rightarrow
    \textcolor{black}{\bigotimes_j^n H_{a=j}\vert00\cdots0\rangle_a\otimes\vert00\cdots0\rangle_t}\nonumber\\
    &=\frac{1}{\sqrt{2^n}}\sum_\mu^{2^n}\vert\mu\rangle_a\otimes\vert00\cdots0\rangle_t,
\end{align}

where $\vert\mu\rangle$ is the $n$ ancillary qubit basis state in the computational basis. Applying a CNOT gate on all target qubits controlled by their respective ancillary qubits yields
\begin{align}
    \vert\Psi\rangle&\rightarrow\bigotimes_j^n U^{\text{CNOT}}_{a=j,t=j}\frac{1}{\sqrt{2^n}}\sum_\mu^{2^n}\vert\mu\rangle_a\otimes\vert00\cdots0\rangle_t\nonumber\\
    &=\frac{1}{\sqrt{2^n}}\sum_\mu^{2^n}\bigotimes_j^n U^{\text{CNOT}}_{a=j,t=j}\vert\mu\rangle_a\otimes\vert00\cdots0\rangle_t\nonumber\\
    &=\frac{1}{\sqrt{2^n}}\sum_\mu^{2^n}\vert\mu\rangle_a\otimes\vert\mu\rangle_t,
\end{align}
\textcolor{black}{where $U^{\text{CNOT}}_{a=j,t=j}$ is a CNOT gate with the $j$th ancillary qubit as control and the $j$th target qubit as target. For simplicity, we can combine the aforementioned Hadamard layer and the CNOT layer as a single operator
\begin{align}
    M &= \bigotimes_j^n U^{\text{CNOT}}_{a=j,t=j} H_{a=j}\\
    \vert\Psi\rangle&=M\vert\Psi\rangle_0.
\end{align}
}
Consider the following probability amplitude:
\begin{equation}
    \langle\Psi\vert U^\dagger V\vert\Psi\rangle=\langle\Psi\vert_0 M U^\dagger V M\vert\Psi\rangle_0.
\end{equation}
This is simply the probability amplitude of measuring the $\vert\Psi\rangle_0$ state after applying the unitary $M U^\dagger V M$ to the initialized $\vert\Psi\rangle_0$ state. Moreover, this probability amplitude is actually the trace of $U^{\dagger}V$:
\begin{align}
    \langle\Psi\vert U^\dagger V\vert\Psi\rangle&=\frac{1}{2^n}\left(\sum_\mu^{2^n}\langle\mu\vert_a\langle\mu\vert_t\right)U^{\dagger}V\left(\sum_\nu^{2^n}\vert\nu\rangle_a\vert\nu\rangle_t\right)\nonumber\\
    &=\frac{1}{2^n}\sum_\mu^{2^n}\langle\mu\vert_t U^{\dagger}V \vert\mu\rangle_t=\frac{1}{2^n}\text{Tr}\left(U^{\dagger}V\right).
\end{align}
Therefore, evaluating the trace becomes evaluating the probability amplitude of obtaining the $\vert\Psi\rangle_0$ state, which QTensor is able to simulate with complexity proportional to the number of qubits and exponential to the circuit depth. This is helpful for evaluating the trace of unitaries that can be efficiently represented by shallow circuits, especially those with limited qubit connectivity such as hardware-efficient ans{\"a}tze.

For qudits with general local dimensions $q$, the generalization is straightforward. We need to replace the Hadamard gate $H$ with the generalized Hadamard gate $H_q$, and the CNOT gate with the $\text{SUM}_q$ gate \cite{qudits}:
\begin{eqnarray}
H_q\vert j\rangle = \frac{1}{\sqrt{q}}\sum_{i=0}^{q-1}e^{2\pi i/q}\vert i\rangle\\
\text{SUM}_q\vert i,j\rangle = \vert i,i+j (\text{mod }q)\rangle.
\end{eqnarray}
Similar to the qubit case, applying the generalized gates to $\vert\Phi\rangle_0$ yields an entangled uniform superposition $\vert\Phi\rangle$ of all basis states. The expectation value of any target qudit unitary with respect to this state is the trace.

\textcolor{black}{Graphically, this can be understood by the tensor equivalence shown in Fig. \ref{trace}b. The gates are
\begin{eqnarray}
H_{ij} = \frac{1}{\sqrt{2}}[\delta_{i 0}+\delta_{i 1}(\delta_{j 0}-\delta_{j 1})]\nonumber\\
U^\text{CNOT}_{kij} = \delta_{k 0}\delta_{i j}+\delta_{k 1 }(\delta_{i 0}\delta_{j 1}+\delta_{i 1}\delta_{j 0}),
\end{eqnarray}
where for $U^\text{CNOT}_{kij}$, $k$ is the control qubit index (CNOT does not change the control qubit in the computational basis and therefore has only one index for the control qubit), and $i,j$ are the target qubit output and input indices. The bottom tensor of Fig. \ref{trace}b evaluates to
\begin{align}
    &H_{m0}U^\text{CNOT}_{mi0}U^\text{CNOT}_{m0j}H_{0m}\nonumber\\
    &
    =\sum_{m}\frac{1}{\sqrt{2}}[\delta_{m 0}+\delta_{m 1}(\delta_{0 0}-\delta_{0 1})]\nonumber\\
    &\times[\delta_{m 0}\delta_{i 0}+\delta_{m 1 }(\delta_{i 0}\delta_{0 1}+\delta_{i 1}\delta_{0 0})]\nonumber\\
    &\times[\delta_{m 0}\delta_{0 j}+\delta_{m 1 }(\delta_{0 0}\delta_{j 1}+\delta_{0 1}\delta_{j 0})]\nonumber\\
    &\times\frac{1}{\sqrt{2}}[\delta_{0 0}+\delta_{0 1}(\delta_{m 0}-\delta_{m 1})]
    =\delta_{i j}.
\end{align}}

\subsection{Verifying the Brown--Susskind conjecture from frame potentials}

 Local and parallel random unitaries are commonly discussed in the context of quantum circuit complexity and the Brown-Susskin conjecture. For both ans{\"a}tzes, the composing random unitaries are drawn from the Haar measure on $U(d^2)$.

\subsubsection{Parallel random unitaries}

\begin{figure} [ht]
   \begin{center}
   \includegraphics[width=7cm]{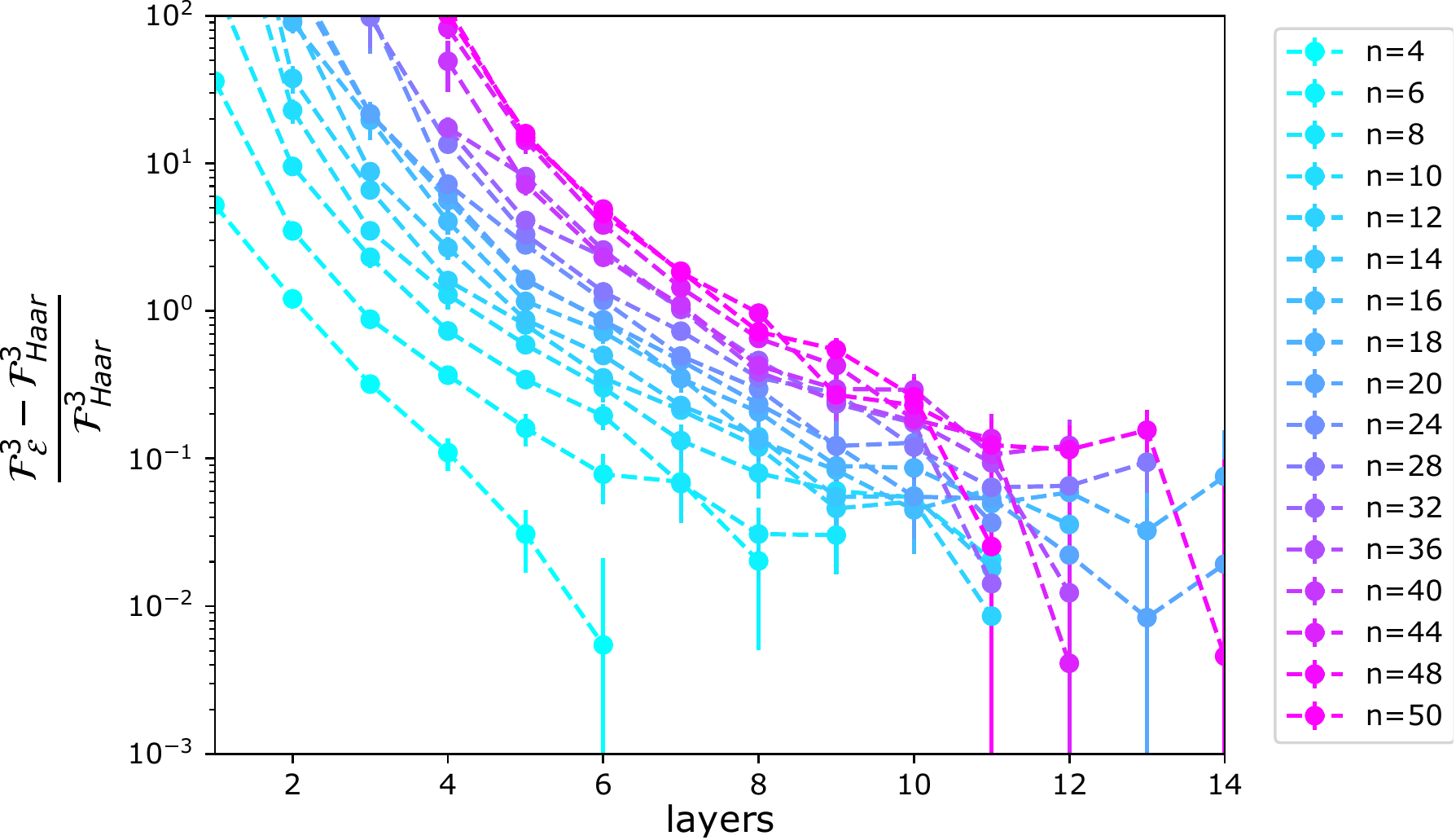}
   \end{center}
   \caption{\textcolor{black}{Fractional} deviation of the $k=3$ frame potential from the Haar value as a function of layers for the parallel random unitary ans{\"a}tze. \textcolor{black}{Error bars correspond to the standard errors.} As shown in Fig. \ref{F theory}, we do not expect linear scaling of $l$ in $n$ with fixed $\mathcal{F}$.}
   { \label{parallel_error}
}
   \end{figure}

\begin{figure} [ht]
   \begin{center}
   \includegraphics[width=7cm]{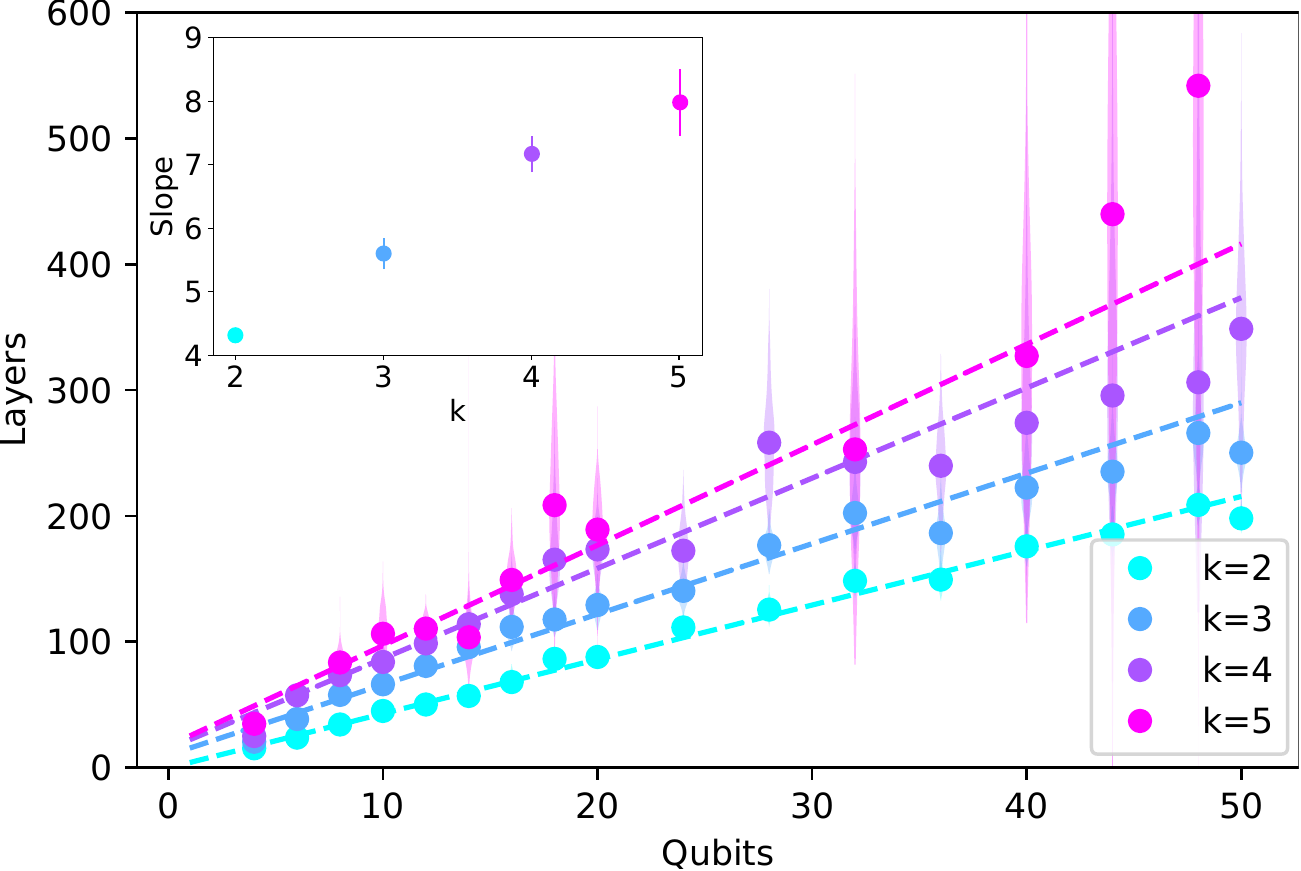}
   \end{center}
   \caption{\textcolor{black}{Layer} scaling as a function of the number of qubits for the parallel random unitary \textcolor{black}{ans{\"a}tze on a violin plot. Solid points are medians of the bootstrap sample, and the vertical shadows represent the sample distribution where the width corresponds to the density. Dotted lines are linear fits.} The inset shows the fitted slopes for different $k$ values. \textcolor{black}{Missing data points are due to insufficient data (see the supplementary materials for more details).}}
   { \label{parallel_fit}
}
   \end{figure}

Results for parallel random unitaries are presented in Figs. \ref{parallel_error} and \ref{parallel_fit}. In Fig. \ref{parallel_error}, The frame potential shows a super-exponential decay in the regime of a few layers and converges to exponential decay as the number of layers increases, just like the theoretical prediction in Fig. \ref{F theory}.

To obtain the layer scaling for reaching $\epsilon$-approximate designs, we fit \textcolor{black}{$\mathcal{F}_\mathcal{E}^{(k)}-\mathcal{F}_\text{Haar}^{(k)}$ to an exponential function according to Eq. \ref{exponential}, and $l$ is estimated using Eq. \ref{layers}.}  Note that our numerical results are in the regime of large $\epsilon$ but we are extrapolating to small $\epsilon$ values, the validity of which depends on a tightly exponentially decaying $\mathcal{F}$.

\textcolor{black}{For robust error analysis, we use bootstrapping to quantify the uncertainties. We randomly sample a subset of computed frame potentials and perform curve fitting to obtain the calculate the number of layers needed to reach $\epsilon<0.1$. This is repeated multiple times to obtain a distribution of layer values, and we use the median of this distribution as our estimate. More details can be found in the supplementary materials.}

\textcolor{black}{Assuming the validity of extrapolation, the results for $\epsilon=0.1$ are shown in Figs. \ref{parallel_fit}. Brandal {\sl et al.} \cite{brandao2016local} established upper bounds on the number of layers needed to approximate $k$-designs for local and parallel random unitaries, which are quadratic and linear in $n$, respectively. They further proved that this bound could not be improved by more than polynomial factors as long as $\epsilon\leq1/4$. Therefore, for us to verify linear growth in $n$, we need to reach under $\epsilon\leq1/4$, which informed our choice of $\epsilon=0.1$ threshold.} We observe a linear scaling of the number of needed layers in $n$, which agrees with the theoretical prediction and non-trivially restricts the $\mathcal{F}$ scale factor $A$ and decay rate $C$, as discussed in Section \ref{frame}. \textcolor{black}{Explanations for missing data points are provided in the supplemental materials.}

Further, we compare the theoretical predictions in Eq. \ref{k=2} against our numerical findings. Figure \ref{parallel_fit} shows the experimental and fitted \textcolor{black}{$k$-design} layer scaling as a function of the number of qubits. Specifically, we fit a linear curve, ignoring the $\log{n}$ and the constant $\log{1/\epsilon}$ terms. We find a slope of \textcolor{black}{$4.38$ in the case of $k=2$}, which is lower than the theoretical value \textcolor{black}{$6.2$ as predicted by E.q. \ref{theory_eq}}. We note,  however, that the theoretical value gives an upper bound of the frame potential since there is overcounting in the contributing domain walls \cite{stat}. Therefore, the analytical expression predicts a larger number of layers needed to approximate 2-designs than necessary. This is apparent in the $n=2$ case, where 16 layers are needed in Eq. \ref{k=2} but a single layer is already sampling from the Haar measure. This accounts for the discrepancy between the theoretical values and the experimental values.

In the inset of Fig. \ref{parallel_fit}, we show the slopes of the scaling curves with different $k$ values. It is predicted that there is a linear $O(nk)$ scaling in $k$ for the number of layers $l$ (or $O(n^2 k)$ scaling for the circuit size $T$) needed to approach $k$-designs \cite{stat}, and a linear relationship between $k$ and complexity is established  in \cite{Brandao:2019sgy}. Together, these findings imply that complexity grows linearly in the circuit size \cite{Brandao:2019sgy, haferkamp2022linear}. Our results support the linear scaling of $T$ in $k$, which predicts that the slope grows linearly in $k$.
  
\subsubsection{Local random unitaries}
  
\begin{figure} [ht]
   \begin{center}
   \includegraphics[width=7cm]{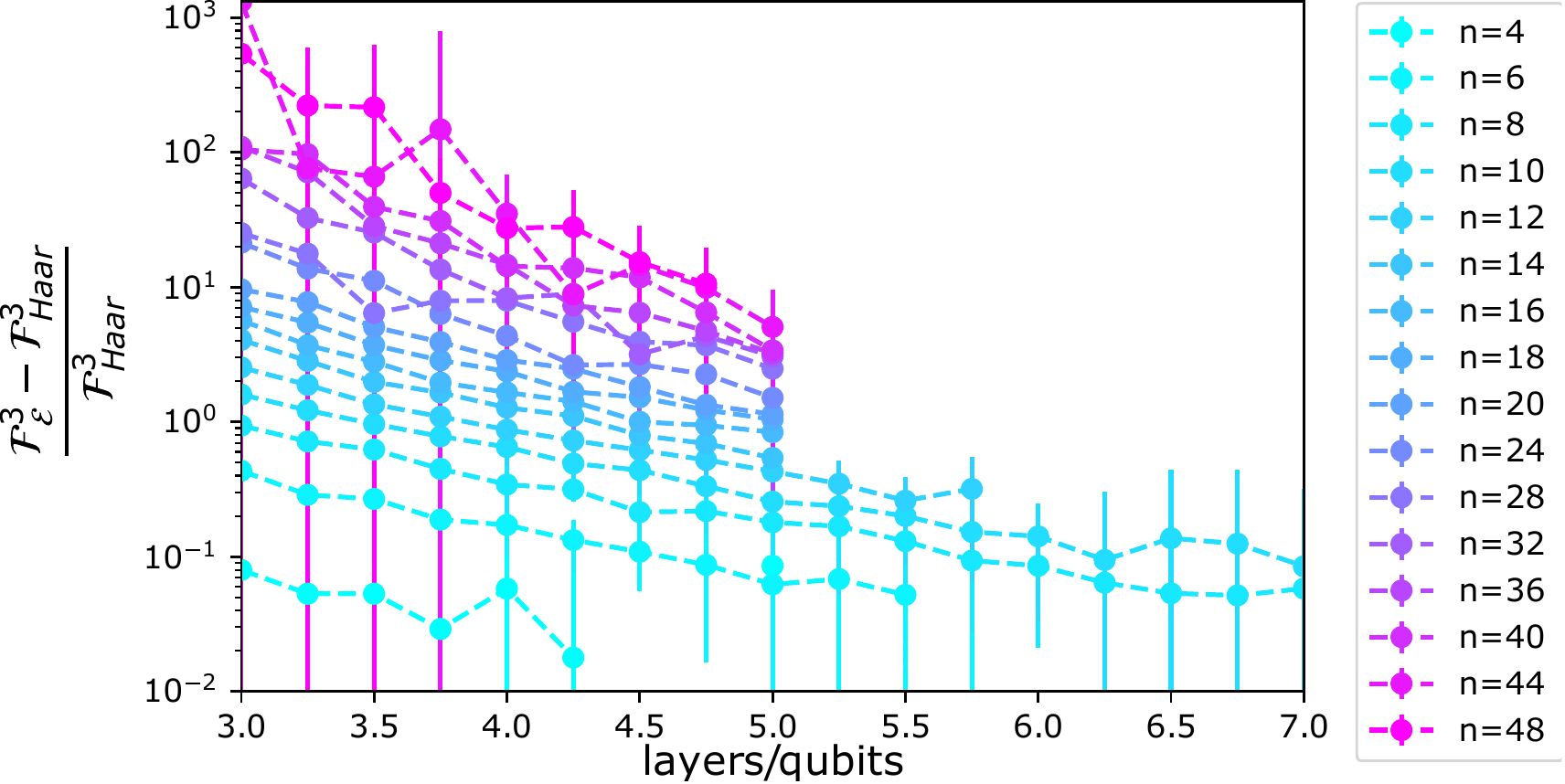}
   \end{center}
   \caption{\textcolor{black}{Fractional} deviation of the $k=3$ frame potential from the Haar value as a function of layers over the number of qubits for the local random unitary ans{\"a}tze.}
   { \label{local_error}
}
   \end{figure}

\begin{figure} [ht]
   \begin{center}
   \includegraphics[width=7cm]{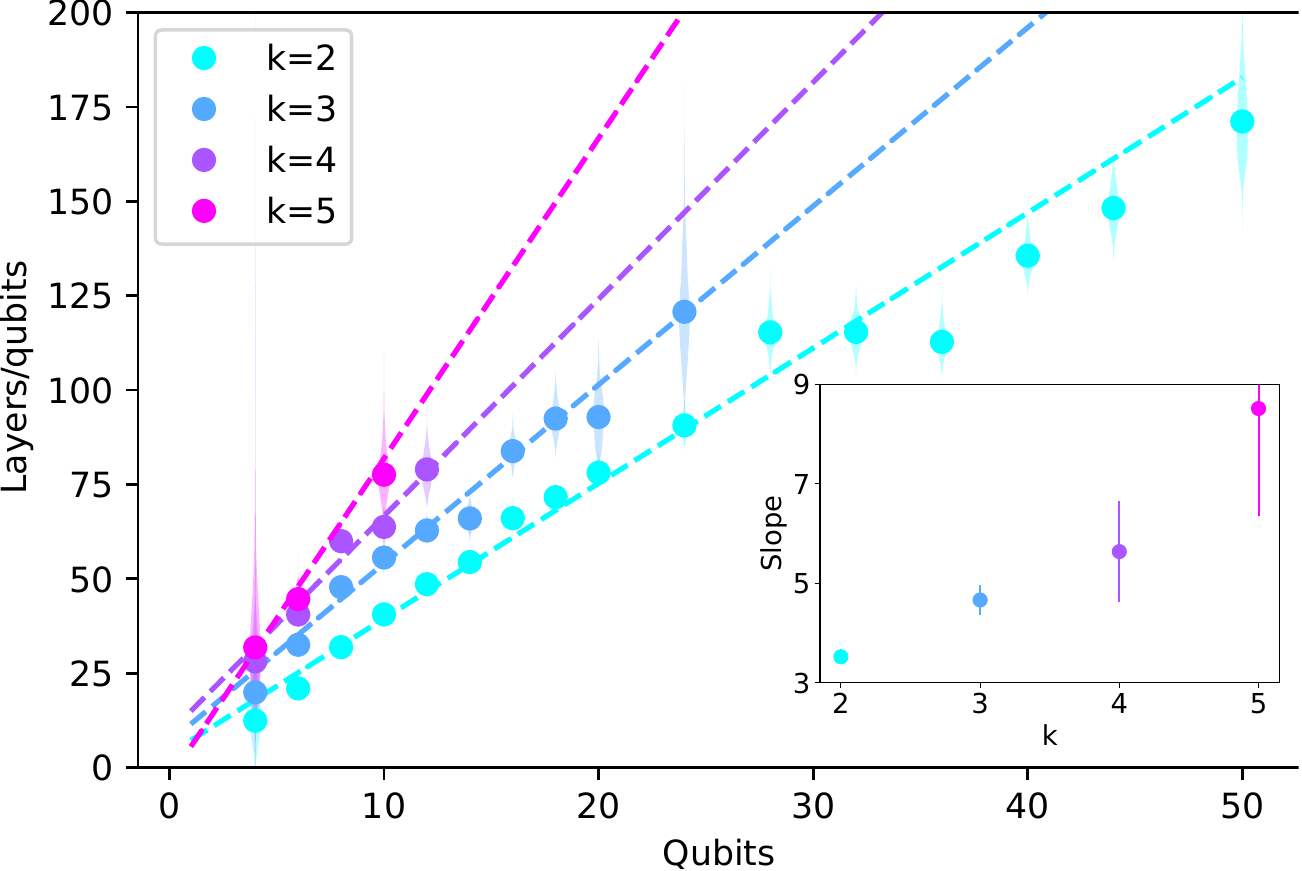}
   \end{center}
   \caption{Layer/qubits scaling as a function of the number of qubits for the local random unitary ans{\"a}tze. The inset shows \textcolor{black}{the fitted slopes for different k values. Missing data points are due to insufficient data (see the supplementary materials for more details).}}
   { \label{local_scaling}
}
   \end{figure}

Results for local random unitaries are presented in Figs. \ref{local_error} and \ref{local_scaling}. Since each layer in the local random circuit has only one gate, we simulate layers proportional to the number of qubits and plot layers/qubits on the $x$-axis to maintain a linear scaling. We observe that this layer/qubits ratio scales linearly with the number of qubits. This is the same gate count scaling as the parallel random unitary ans{\"a}tze, both quadratic in $n$. The scaling in $k$ is \textcolor{black}{close to linear, but the confidence is lower due to a lack of data points for $k=4,5$ at large $n$. Explanations for missing data points are provided in the supplemental materials.}

\subsection{Hardware-efficient ans{\"a}tze as approximate $k$-designs}

\begin{figure} [ht]
   \begin{center}
   \includegraphics[width=7cm]{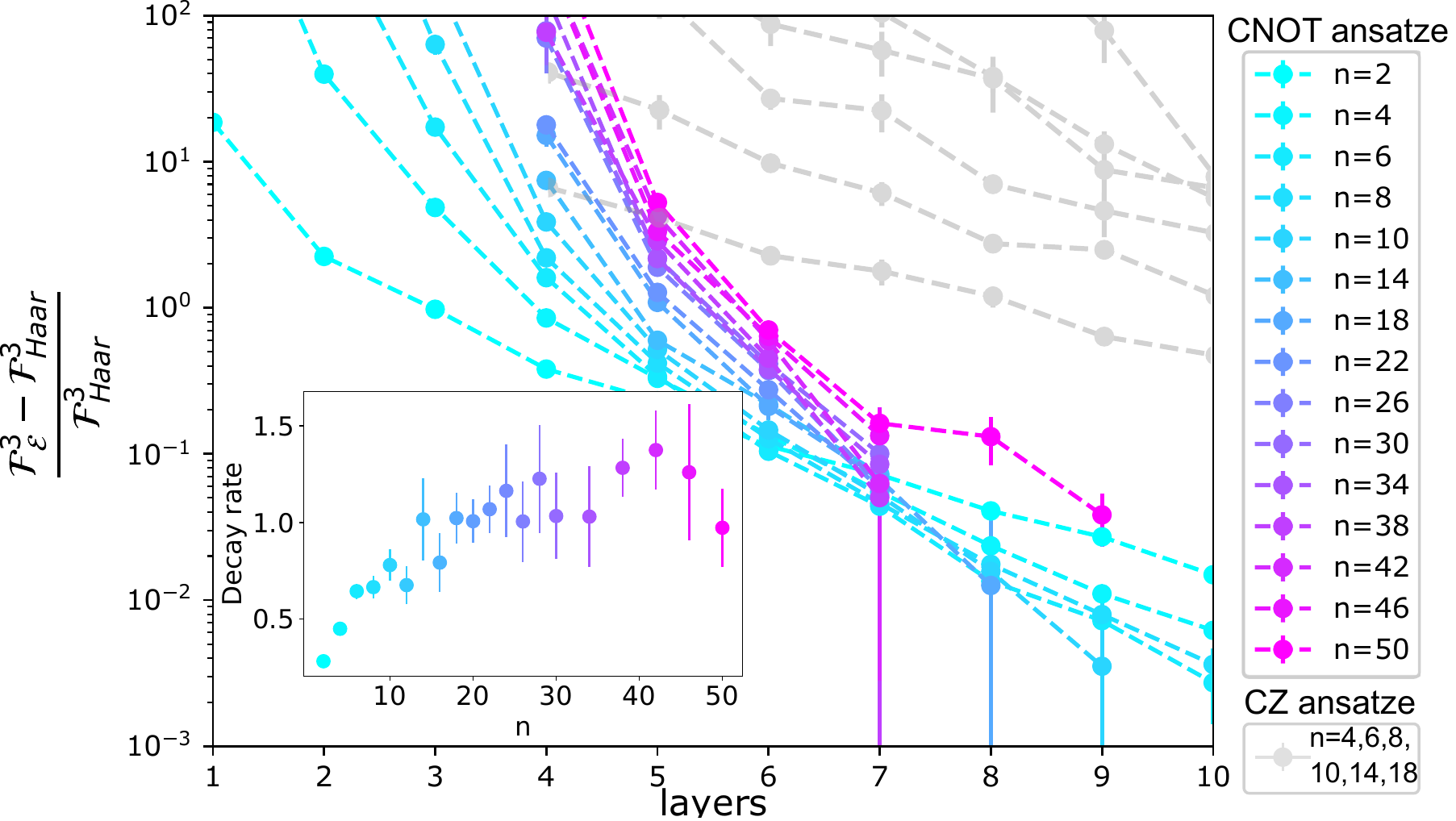}
   \end{center}
   \caption{\textcolor{black}{Fractional} deviation of the $k=3$ frame potential from the Haar value as a function of layers for the hardware-efficient ans{\"a}tze. \textcolor{black}{Colorful traces are for the CNOT gate-based ans{\"a}tze, and gray traces are for the CZ gate-based ans{\"a}tze.} The inset shows the \textcolor{black}{CNOT ans{\"a}tze} decay rate scaling of $\mathcal{F}$ in the number of qubits $n$.}
   { \label{hardware_efficient_error}
}
   \end{figure}

\begin{figure} [ht]
   \begin{center}
   \includegraphics[width=7cm]{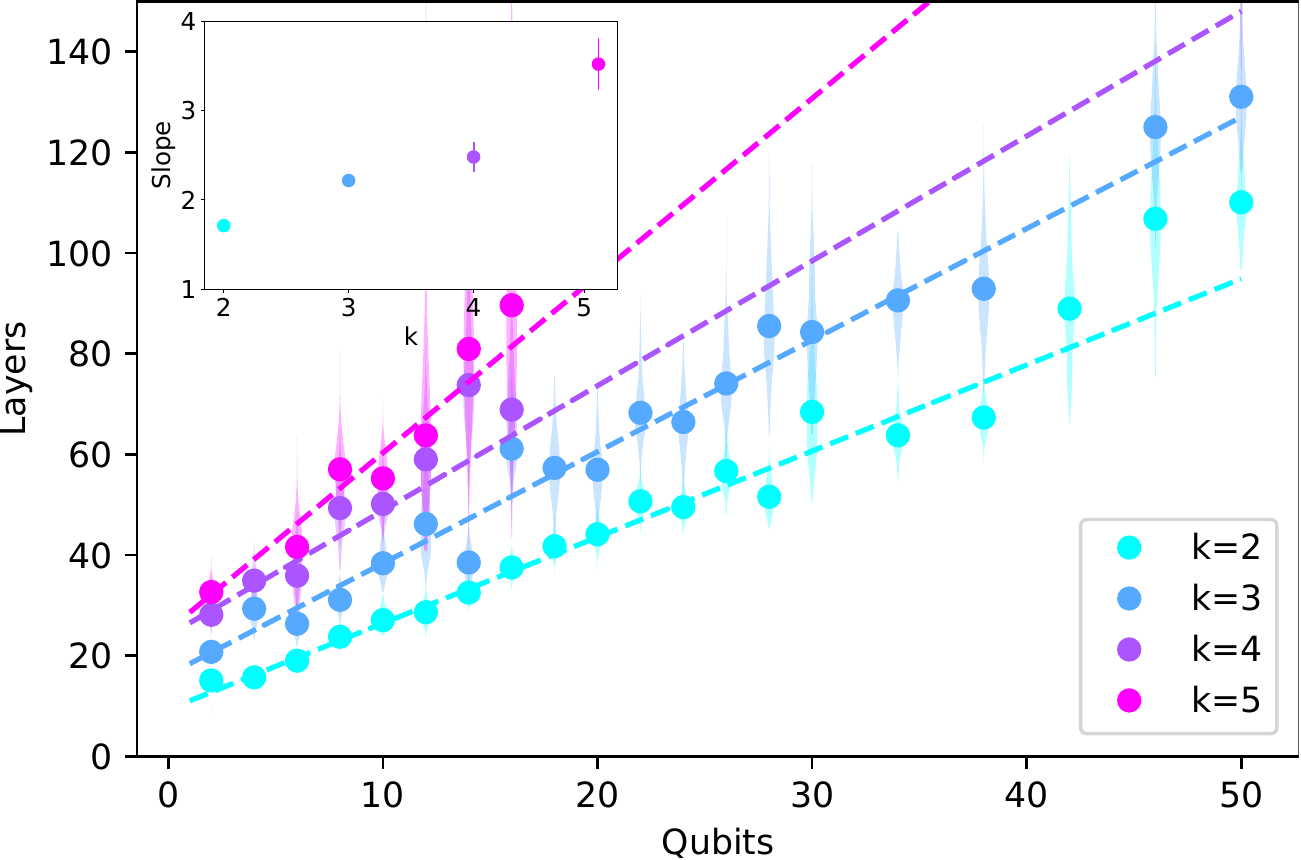}
   \end{center}
   \caption{Layer scaling as a function of the number of qubits for the hardware-efficient ans{\"a}tze. \textcolor{black}{The inset shows the fitted slopes for different k values. Missing data points are due to insufficient data (see supplementary materials for more details).}}
   { \label{hardware_efficient_scaling}
}
   \end{figure}

Originally proposed as  ans{\"a}tze for variational quantum eigensolvers \cite{kandala2017hardware}, hardware-efficient ans{\"a}tze utilize gates and connectivity readily available on the quantum hardware and are attractive because of their relaxed hardware requirements \textcolor{black}{\cite{grossi2022finite,nakaji2021expressibility,Du2022search}}. In addition, the ans{\"a}tze are simulated in the context of the barren plateau problem \cite{mcclean2018barren}, \textcolor{black}{where the variance of gradients vanish exponentially with the number of qubits in sufficiently deep circuits. In fact, the proof of the barren plateau problem assumes that circuits before and after the gate whose derivative we are computing are approximate 2-designs, which is especially suitable for the hardware efficient ans{\"a}tze because they are believed to be efficient at scrambling. We simulated these circuits with controlled-phased gates and controlled-not gates as two-qubit gates, respectively. Figure \ref{hardware_efficient_error} shows that the controlled-not gate based ans{\"a}tzee approaches the Haar measure sooner, and therefore future analysis is conducted on CNOT based ans{\"a}tze only.} Figure \ref{hardware_efficient_scaling} shows a linear dependence on the number of qubits, as well as \textcolor{black}{a positive dependence on $k$. Explanations for missing data points are provided in the supplemental materials.}

We note that these ans{\"a}tze reach lower frame potential values with much fewer layers, albeit having much fewer parameters per layer. This result is partially explainable through the observation that each layer in the hardware-efficient ans{\"a}tze contains two layers of two-qubit gate walls, whereas each layer in the parallel random unitary ans{\"a}tze contains only one wall. Further, random unitaries from $U(d^2)$ are not all maximally entangling. The hardware-efficient ans{\"a}tze can therefore generate highly entangled stages much more efficiently, exploring a much larger space with fewer parameters.

Further, unlike the previously discussed ans{\"a}tze where the frame potential decay rate is constant, the hardware-efficient ans{\"a}tze decay rate increases with $n$ as shown in the inset of Fig. \ref{hardware_efficient_error}. This does not contradict the observed linear scaling as long as the decay rate scaling is sublinear.

This observation confirms that hardware-efficient ans{\"a}tze are highly expressive, a concept that is crucial to the utility of variational quantum algorithms. Ans{\"a}tze with higher expressibility are able to better represent the Haar distribution and, therefore able to better approximate the target unitary or minimize the objective. This links the expressibility to the frame potential \cite{expressibility}. The high expressibility of hardware-efficient ans{\"a}tze and their close relatives, and consequently the desirable noise properties due to their shallow depths, are precisely the argument in favor of these ans{\"a}tze over their deeper and more complex problem-aware counterparts \cite{liu2022layer}. With the recent discovery of the relation between expressibility and gradient variance \cite{expressibility_barren}, the analysis of frame potentials can play an important role in theoretically and empirically determining the usefulness of various ans{\"a}tze for variational algorithms.

\section{Discussion}\label{conclude}

Evaluating the distance from a given random circuit ensemble to the exact Haar randomness is important for understanding several perspectives in quantum information science, including recent experiments on the near-term quantum advantage. Explicitly constructing the unitaries requires memory complexity of $O(4^{n})$. A more efficient classical algorithm decomposes a unitary into gates in a universal set ($H$, $T$, and CNOT), which allows us to estimate the normalized trace by sampling allowed Feynman paths \cite{dqc1}. Exact evaluation using this method is NP-complete, and approximation to fixed precision requires a number of Feynman path samples that are exponentially large in the number of Hadamard gates in the circuit. Fortunately, for shallow circuits, tensor-network-based algorithms can obtain the \textit{exact} trace with \textit{linear} complexity in $n$.

In our paper, we simulate large-scale random circuit sampling experiments classically up to 50 qubits, the number of noisy physical qubits we are able to control in the NISQ era, using the \texttt{QTensor} package. As examples, we provide two applications of our computational tools: a numerical verification of the Brown--Susskind conjecture and a numerical estimation relating to the barren plateau in quantum machine learning and the randomized hardware-efficient variational ans{\"a}tze.

Through our examples, we show that classical tensor network simulations are useful for our understanding of open problems in theoretical computer science and numerical examinations of quantum neural network properties for quantum computing applications. We believe that tensor networks and other cutting-edge tools are useful for probing the boundary of classical simulation and improving the understanding of quantum advantage in several subjects of quantum physics, for instance, quantum simulation \cite{Yuan:2020xmq,Milsted:2020jmf}. Moreover, it will  be interesting to connect our algorithms to the current research on classical simulation of boson sampling experiments.

\section{Methods}\label{evaframe}

\subsection{Tensor network simulator}

For all the trace evaluations, we use the QTensorAI library \cite{liu2022embedding}, originally developed to simulate quantum machine learning with parameterized circuits. This library allows quantum circuits to be simulated in parallel on CPUs and GPUs, which is a highly desirable property for sampling a large number of circuits. The library is based on the QTensor simulator \cite{lykov2021performance,lykov_diagonal,lykov2021large}, a tensor network-based quantum simulator that represents the network as an undirected graph.

In this method of simulation, the computation is memory bound, and the memory complexity is exponential in the ``treewidth,'' the largest rank of tensor that needs to be stored during computation. The graphical formalism utilized by QTensor allows the tensor contraction order to be optimized to minimize the treewidth. For shallow quantum circuits, the treewidth is determined mainly  by the number of layers in the quantum circuit, and therefore QTensor is particularly well suited for simulating shallow circuits such as those used in the Quantum Approximate Optimization Algorithm (QAOA).

\subsection{Sampling $U(d^2)$}

The simulation of both parallel and local random unitary circuits requires the use of random two-qubit random unitary gates. We implement these gates and sample Haar unitaries according to the scheme proposed for unitary neural networks \cite{jing2017tunable}, using a PyTorch implementation \cite{eunn}. \textcolor{black}{The universality of this decomposition scheme is first proved in the context of optical interferometers \cite{Reck1994, Clements2016}.} This implementation parameterizes two-qubit unitaries using 16 phase parameters, and uniformly sampling these parameters leads to uniform sampling on the Haar measure. Further, it is fully differentiable, although we do not care about this property in this work.

\subsection{High-performance computing}

For hardware-efficient and parallel random unitary ans{\"a}tze, once the number of qubits and the number of layers are chosen, the circuit topology will remain the same throughout the ensemble. This is in contrast to the local random unitary ans{\"a}tze, where a two-qubit gate is applied to random neighboring qubits in each layer, which means that the circuit topologies are very different within an ensemble. For fixed-topology ensembles, the algorithm can optimize the contraction order for all circuits at once. This optimization significantly reduces the computational complexity, and the optimization time is on the order of minutes depending on the circuit size. However, local random unitary circuits cannot benefit from circuit optimizations since we would need to do that for each sample, whereas the actual simulation time is usually much shorter.

Further, for fixed topology circuits, the tensor contraction operations are identical, which is very suitable for single-instruction multi-data parallel executions on GPUs. For ensembles with the smallest tree widths, we can compute the trace values of millions of circuits in parallel on a single GPU. However, local random unitary circuits are not compatible with single-instruction parallel computation and must be simulated in parallel using a CPU cluster.

\section*{Data availability}
 Data containing the bootstrap frame potential values used to generate the figures are available in the \textcolor{black}{GitHub repository \cite{fp}}, and data for the calculated trace values of sampled random circuits is available upon request from the authors.

\section*{Code availability}
The code used to generate the data and figures is available in the GitHub repository \cite{fp}. The tensor network quantum simulator QTensor and QTensorAI are open source, and available \cite{qtensor_ai, qtensor}.

\bibliography{frame.bib}

\begin{thebibliography}{83}%
\makeatletter
\providecommand \@ifxundefined [1]{%
 \@ifx{#1\undefined}
}%
\providecommand \@ifnum [1]{%
 \ifnum #1\expandafter \@firstoftwo
 \else \expandafter \@secondoftwo
 \fi
}%
\providecommand \@ifx [1]{%
 \ifx #1\expandafter \@firstoftwo
 \else \expandafter \@secondoftwo
 \fi
}%
\providecommand \natexlab [1]{#1}%
\providecommand \enquote  [1]{``#1''}%
\providecommand \bibnamefont  [1]{#1}%
\providecommand \bibfnamefont [1]{#1}%
\providecommand \citenamefont [1]{#1}%
\providecommand \href@noop [0]{\@secondoftwo}%
\providecommand \href [0]{\begingroup \@sanitize@url \@href}%
\providecommand \@href[1]{\@@startlink{#1}\@@href}%
\providecommand \@@href[1]{\endgroup#1\@@endlink}%
\providecommand \@sanitize@url [0]{\catcode `\\12\catcode `\$12\catcode
  `\&12\catcode `\#12\catcode `\^12\catcode `\_12\catcode `\%12\relax}%
\providecommand \@@startlink[1]{}%
\providecommand \@@endlink[0]{}%
\providecommand \url  [0]{\begingroup\@sanitize@url \@url }%
\providecommand \@url [1]{\endgroup\@href {#1}{\urlprefix }}%
\providecommand \urlprefix  [0]{URL }%
\providecommand \Eprint [0]{\href }%
\providecommand \doibase [0]{http://dx.doi.org/}%
\providecommand \selectlanguage [0]{\@gobble}%
\providecommand \bibinfo  [0]{\@secondoftwo}%
\providecommand \bibfield  [0]{\@secondoftwo}%
\providecommand \translation [1]{[#1]}%
\providecommand \BibitemOpen [0]{}%
\providecommand \bibitemStop [0]{}%
\providecommand \bibitemNoStop [0]{.\EOS\space}%
\providecommand \EOS [0]{\spacefactor3000\relax}%
\providecommand \BibitemShut  [1]{\csname bibitem#1\endcsname}%
\let\auto@bib@innerbib\@empty
\bibitem [{\citenamefont {Feynman}\ \emph {et~al.}(1982)\citenamefont {Feynman}
  \emph {et~al.}}]{feynman1982simulating}%
  \BibitemOpen
  \bibfield  {author} {\bibinfo {author} {\bibfnamefont {R.~P.}\ \bibnamefont
  {Feynman}} \emph {et~al.},\ }\href@noop {} {\bibfield  {journal} {\bibinfo
  {journal} {Int. j. Theor. Phys}\ }\textbf {\bibinfo {volume} {21}} (\bibinfo
  {year} {1982})}\BibitemShut {NoStop}%
\bibitem [{\citenamefont {Preskill}(2012)}]{preskill2012quantum}%
  \BibitemOpen
  \bibfield  {author} {\bibinfo {author} {\bibfnamefont {J.}~\bibnamefont
  {Preskill}},\ }\href@noop {} {\  (\bibinfo {year} {2012})},\ \bibinfo {note}
  {preprint at \url{https://arxiv.org/abs/1203.5813}}\BibitemShut {NoStop}%
\bibitem [{\citenamefont {Alexeev}\ \emph {et~al.}(2021)\citenamefont
  {Alexeev}, \citenamefont {Bacon}, \citenamefont {Brown}, \citenamefont
  {Calderbank}, \citenamefont {Carr}, \citenamefont {Chong}, \citenamefont
  {DeMarco}, \citenamefont {Englund}, \citenamefont {Farhi}, \citenamefont
  {Fefferman} \emph {et~al.}}]{alexeev2021quantum}%
  \BibitemOpen
  \bibfield  {author} {\bibinfo {author} {\bibfnamefont {Y.}~\bibnamefont
  {Alexeev}}, \bibinfo {author} {\bibfnamefont {D.}~\bibnamefont {Bacon}},
  \bibinfo {author} {\bibfnamefont {K.~R.}\ \bibnamefont {Brown}}, \bibinfo
  {author} {\bibfnamefont {R.}~\bibnamefont {Calderbank}}, \bibinfo {author}
  {\bibfnamefont {L.~D.}\ \bibnamefont {Carr}}, \bibinfo {author}
  {\bibfnamefont {F.~T.}\ \bibnamefont {Chong}}, \bibinfo {author}
  {\bibfnamefont {B.}~\bibnamefont {DeMarco}}, \bibinfo {author} {\bibfnamefont
  {D.}~\bibnamefont {Englund}}, \bibinfo {author} {\bibfnamefont
  {E.}~\bibnamefont {Farhi}}, \bibinfo {author} {\bibfnamefont
  {B.}~\bibnamefont {Fefferman}},  \emph {et~al.},\ }\href@noop {} {\bibfield
  {journal} {\bibinfo  {journal} {PRX Quantum}\ }\textbf {\bibinfo {volume}
  {2}},\ \bibinfo {pages} {017001} (\bibinfo {year} {2021})}\BibitemShut
  {NoStop}%
\bibitem [{\citenamefont {Preskill}(2018)}]{preskill2018quantum}%
  \BibitemOpen
  \bibfield  {author} {\bibinfo {author} {\bibfnamefont {J.}~\bibnamefont
  {Preskill}},\ }\href@noop {} {\bibfield  {journal} {\bibinfo  {journal}
  {Quantum}\ }\textbf {\bibinfo {volume} {2}},\ \bibinfo {pages} {79} (\bibinfo
  {year} {2018})}\BibitemShut {NoStop}%
\bibitem [{\citenamefont {Arute}\ \emph {et~al.}(2019)\citenamefont {Arute},
  \citenamefont {Arya}, \citenamefont {Babbush}, \citenamefont {Bacon},
  \citenamefont {Bardin}, \citenamefont {Barends}, \citenamefont {Biswas},
  \citenamefont {Boixo}, \citenamefont {Brandao}, \citenamefont {Buell} \emph
  {et~al.}}]{arute2019quantum}%
  \BibitemOpen
  \bibfield  {author} {\bibinfo {author} {\bibfnamefont {F.}~\bibnamefont
  {Arute}}, \bibinfo {author} {\bibfnamefont {K.}~\bibnamefont {Arya}},
  \bibinfo {author} {\bibfnamefont {R.}~\bibnamefont {Babbush}}, \bibinfo
  {author} {\bibfnamefont {D.}~\bibnamefont {Bacon}}, \bibinfo {author}
  {\bibfnamefont {J.~C.}\ \bibnamefont {Bardin}}, \bibinfo {author}
  {\bibfnamefont {R.}~\bibnamefont {Barends}}, \bibinfo {author} {\bibfnamefont
  {R.}~\bibnamefont {Biswas}}, \bibinfo {author} {\bibfnamefont
  {S.}~\bibnamefont {Boixo}}, \bibinfo {author} {\bibfnamefont {F.~G.}\
  \bibnamefont {Brandao}}, \bibinfo {author} {\bibfnamefont {D.~A.}\
  \bibnamefont {Buell}},  \emph {et~al.},\ }\href@noop {} {\bibfield  {journal}
  {\bibinfo  {journal} {Nature}\ }\textbf {\bibinfo {volume} {574}},\ \bibinfo
  {pages} {505} (\bibinfo {year} {2019})}\BibitemShut {NoStop}%
\bibitem [{\citenamefont {Zhong}\ \emph {et~al.}(2020)\citenamefont {Zhong},
  \citenamefont {Wang}, \citenamefont {Deng}, \citenamefont {Chen},
  \citenamefont {Peng}, \citenamefont {Luo}, \citenamefont {Qin}, \citenamefont
  {Wu}, \citenamefont {Ding}, \citenamefont {Hu} \emph
  {et~al.}}]{zhong2020quantum}%
  \BibitemOpen
  \bibfield  {author} {\bibinfo {author} {\bibfnamefont {H.-S.}\ \bibnamefont
  {Zhong}}, \bibinfo {author} {\bibfnamefont {H.}~\bibnamefont {Wang}},
  \bibinfo {author} {\bibfnamefont {Y.-H.}\ \bibnamefont {Deng}}, \bibinfo
  {author} {\bibfnamefont {M.-C.}\ \bibnamefont {Chen}}, \bibinfo {author}
  {\bibfnamefont {L.-C.}\ \bibnamefont {Peng}}, \bibinfo {author}
  {\bibfnamefont {Y.-H.}\ \bibnamefont {Luo}}, \bibinfo {author} {\bibfnamefont
  {J.}~\bibnamefont {Qin}}, \bibinfo {author} {\bibfnamefont {D.}~\bibnamefont
  {Wu}}, \bibinfo {author} {\bibfnamefont {X.}~\bibnamefont {Ding}}, \bibinfo
  {author} {\bibfnamefont {Y.}~\bibnamefont {Hu}},  \emph {et~al.},\
  }\href@noop {} {\bibfield  {journal} {\bibinfo  {journal} {Science}\ }\textbf
  {\bibinfo {volume} {370}},\ \bibinfo {pages} {1460} (\bibinfo {year}
  {2020})}\BibitemShut {NoStop}%
\bibitem [{\citenamefont {Ebadi}\ \emph {et~al.}(2022)\citenamefont {Ebadi},
  \citenamefont {Keesling}, \citenamefont {Cain}, \citenamefont {Wang},
  \citenamefont {Levine}, \citenamefont {Bluvstein}, \citenamefont {Semeghini},
  \citenamefont {Omran}, \citenamefont {Liu}, \citenamefont {Samajdar},
  \citenamefont {Luo}, \citenamefont {Nash}, \citenamefont {Gao}, \citenamefont
  {Barak}, \citenamefont {Farhi}, \citenamefont {Sachdev}, \citenamefont
  {Gemelke}, \citenamefont {Zhou}, \citenamefont {Choi}, \citenamefont
  {Pichler}, \citenamefont {Wang}, \citenamefont {Greiner}, \citenamefont
  {Vuletic},\ and\ \citenamefont {Lukin}}]{ebadi2022quantum}%
  \BibitemOpen
  \bibfield  {author} {\bibinfo {author} {\bibfnamefont {S.}~\bibnamefont
  {Ebadi}}, \bibinfo {author} {\bibfnamefont {A.}~\bibnamefont {Keesling}},
  \bibinfo {author} {\bibfnamefont {M.}~\bibnamefont {Cain}}, \bibinfo {author}
  {\bibfnamefont {T.~T.}\ \bibnamefont {Wang}}, \bibinfo {author}
  {\bibfnamefont {H.}~\bibnamefont {Levine}}, \bibinfo {author} {\bibfnamefont
  {D.}~\bibnamefont {Bluvstein}}, \bibinfo {author} {\bibfnamefont
  {G.}~\bibnamefont {Semeghini}}, \bibinfo {author} {\bibfnamefont
  {A.}~\bibnamefont {Omran}}, \bibinfo {author} {\bibfnamefont
  {J.}~\bibnamefont {Liu}}, \bibinfo {author} {\bibfnamefont {R.}~\bibnamefont
  {Samajdar}}, \bibinfo {author} {\bibfnamefont {X.-Z.}\ \bibnamefont {Luo}},
  \bibinfo {author} {\bibfnamefont {B.}~\bibnamefont {Nash}}, \bibinfo {author}
  {\bibfnamefont {X.}~\bibnamefont {Gao}}, \bibinfo {author} {\bibfnamefont
  {B.}~\bibnamefont {Barak}}, \bibinfo {author} {\bibfnamefont
  {E.}~\bibnamefont {Farhi}}, \bibinfo {author} {\bibfnamefont
  {S.}~\bibnamefont {Sachdev}}, \bibinfo {author} {\bibfnamefont
  {N.}~\bibnamefont {Gemelke}}, \bibinfo {author} {\bibfnamefont
  {L.}~\bibnamefont {Zhou}}, \bibinfo {author} {\bibfnamefont {S.}~\bibnamefont
  {Choi}}, \bibinfo {author} {\bibfnamefont {H.}~\bibnamefont {Pichler}},
  \bibinfo {author} {\bibfnamefont {S.}~\bibnamefont {Wang}}, \bibinfo {author}
  {\bibfnamefont {M.}~\bibnamefont {Greiner}}, \bibinfo {author} {\bibfnamefont
  {V.}~\bibnamefont {Vuletic}}, \ and\ \bibinfo {author} {\bibfnamefont
  {M.~D.}\ \bibnamefont {Lukin}},\ }\href@noop {} {\enquote {\bibinfo {title}
  {Quantum optimization of maximum independent set using {Rydberg} atom
  arrays},}\ } (\bibinfo {year} {2022}),\ \bibinfo {note} {preprint at
  \url{https://arxiv.org/abs/2202.09372}}\BibitemShut {NoStop}%
\bibitem [{\citenamefont {White}(1992)}]{white1992density}%
  \BibitemOpen
  \bibfield  {author} {\bibinfo {author} {\bibfnamefont {S.~R.}\ \bibnamefont
  {White}},\ }\href@noop {} {\bibfield  {journal} {\bibinfo  {journal}
  {Physical Review Letters}\ }\textbf {\bibinfo {volume} {69}},\ \bibinfo
  {pages} {2863} (\bibinfo {year} {1992})}\BibitemShut {NoStop}%
\bibitem [{\citenamefont {Rommer}\ and\ \citenamefont
  {Ostlund}(1997)}]{rommer1997class}%
  \BibitemOpen
  \bibfield  {author} {\bibinfo {author} {\bibfnamefont {S.}~\bibnamefont
  {Rommer}}\ and\ \bibinfo {author} {\bibfnamefont {S.}~\bibnamefont
  {Ostlund}},\ }\href@noop {} {\bibfield  {journal} {\bibinfo  {journal}
  {Physical Review B}\ }\textbf {\bibinfo {volume} {55}},\ \bibinfo {pages}
  {2164} (\bibinfo {year} {1997})}\BibitemShut {NoStop}%
\bibitem [{\citenamefont {Orus}(2019)}]{orus2019tensor}%
  \BibitemOpen
  \bibfield  {author} {\bibinfo {author} {\bibfnamefont {R.}~\bibnamefont
  {Orus}},\ }\href@noop {} {\bibfield  {journal} {\bibinfo  {journal} {Nature
  Reviews Physics}\ }\textbf {\bibinfo {volume} {1}},\ \bibinfo {pages} {538}
  (\bibinfo {year} {2019})}\BibitemShut {NoStop}%
\bibitem [{\citenamefont {Noh}\ \emph {et~al.}(2020)\citenamefont {Noh},
  \citenamefont {Jiang},\ and\ \citenamefont {Fefferman}}]{noh2020efficient}%
  \BibitemOpen
  \bibfield  {author} {\bibinfo {author} {\bibfnamefont {K.}~\bibnamefont
  {Noh}}, \bibinfo {author} {\bibfnamefont {L.}~\bibnamefont {Jiang}}, \ and\
  \bibinfo {author} {\bibfnamefont {B.}~\bibnamefont {Fefferman}},\ }\href@noop
  {} {\bibfield  {journal} {\bibinfo  {journal} {Quantum}\ }\textbf {\bibinfo
  {volume} {4}},\ \bibinfo {pages} {318} (\bibinfo {year} {2020})}\BibitemShut
  {NoStop}%
\bibitem [{\citenamefont {Huang}\ \emph {et~al.}(2020)\citenamefont {Huang},
  \citenamefont {Zhang}, \citenamefont {Newman}, \citenamefont {Cai},
  \citenamefont {Gao}, \citenamefont {Tian}, \citenamefont {Wu}, \citenamefont
  {Xu}, \citenamefont {Yu}, \citenamefont {Yuan} \emph
  {et~al.}}]{huang2020classical}%
  \BibitemOpen
  \bibfield  {author} {\bibinfo {author} {\bibfnamefont {C.}~\bibnamefont
  {Huang}}, \bibinfo {author} {\bibfnamefont {F.}~\bibnamefont {Zhang}},
  \bibinfo {author} {\bibfnamefont {M.}~\bibnamefont {Newman}}, \bibinfo
  {author} {\bibfnamefont {J.}~\bibnamefont {Cai}}, \bibinfo {author}
  {\bibfnamefont {X.}~\bibnamefont {Gao}}, \bibinfo {author} {\bibfnamefont
  {Z.}~\bibnamefont {Tian}}, \bibinfo {author} {\bibfnamefont {J.}~\bibnamefont
  {Wu}}, \bibinfo {author} {\bibfnamefont {H.}~\bibnamefont {Xu}}, \bibinfo
  {author} {\bibfnamefont {H.}~\bibnamefont {Yu}}, \bibinfo {author}
  {\bibfnamefont {B.}~\bibnamefont {Yuan}},  \emph {et~al.},\ }\href@noop {} {\
   (\bibinfo {year} {2020})},\ \bibinfo {note} {preprint at
  \url{https://arxiv.org/abs/2005.06787}}\BibitemShut {NoStop}%
\bibitem [{\citenamefont {Pan}\ and\ \citenamefont
  {Zhang}(2021)}]{pan2021simulating}%
  \BibitemOpen
  \bibfield  {author} {\bibinfo {author} {\bibfnamefont {F.}~\bibnamefont
  {Pan}}\ and\ \bibinfo {author} {\bibfnamefont {P.}~\bibnamefont {Zhang}},\
  }\href@noop {} {\  (\bibinfo {year} {2021})},\ \bibinfo {note} {preprint at
  \url{https://arxiv.org/abs/2103.03074}}\BibitemShut {NoStop}%
\bibitem [{\citenamefont {Oh}\ \emph {et~al.}(2021)\citenamefont {Oh},
  \citenamefont {Noh}, \citenamefont {Fefferman},\ and\ \citenamefont
  {Jiang}}]{oh2021classical}%
  \BibitemOpen
  \bibfield  {author} {\bibinfo {author} {\bibfnamefont {C.}~\bibnamefont
  {Oh}}, \bibinfo {author} {\bibfnamefont {K.}~\bibnamefont {Noh}}, \bibinfo
  {author} {\bibfnamefont {B.}~\bibnamefont {Fefferman}}, \ and\ \bibinfo
  {author} {\bibfnamefont {L.}~\bibnamefont {Jiang}},\ }\href@noop {}
  {\bibfield  {journal} {\bibinfo  {journal} {Physical Review A}\ }\textbf
  {\bibinfo {volume} {104}},\ \bibinfo {pages} {022407} (\bibinfo {year}
  {2021})}\BibitemShut {NoStop}%
\bibitem [{\citenamefont {Boixo}\ \emph {et~al.}(2018)\citenamefont {Boixo},
  \citenamefont {Isakov}, \citenamefont {Smelyanskiy}, \citenamefont {Babbush},
  \citenamefont {Ding}, \citenamefont {Jiang}, \citenamefont {Bremner},
  \citenamefont {Martinis},\ and\ \citenamefont
  {Neven}}]{boixo2018characterizing}%
  \BibitemOpen
  \bibfield  {author} {\bibinfo {author} {\bibfnamefont {S.}~\bibnamefont
  {Boixo}}, \bibinfo {author} {\bibfnamefont {S.~V.}\ \bibnamefont {Isakov}},
  \bibinfo {author} {\bibfnamefont {V.~N.}\ \bibnamefont {Smelyanskiy}},
  \bibinfo {author} {\bibfnamefont {R.}~\bibnamefont {Babbush}}, \bibinfo
  {author} {\bibfnamefont {N.}~\bibnamefont {Ding}}, \bibinfo {author}
  {\bibfnamefont {Z.}~\bibnamefont {Jiang}}, \bibinfo {author} {\bibfnamefont
  {M.~J.}\ \bibnamefont {Bremner}}, \bibinfo {author} {\bibfnamefont {J.~M.}\
  \bibnamefont {Martinis}}, \ and\ \bibinfo {author} {\bibfnamefont
  {H.}~\bibnamefont {Neven}},\ }\href@noop {} {\bibfield  {journal} {\bibinfo
  {journal} {Nature Physics}\ }\textbf {\bibinfo {volume} {14}},\ \bibinfo
  {pages} {595} (\bibinfo {year} {2018})}\BibitemShut {NoStop}%
\bibitem [{\citenamefont {Harrow}\ and\ \citenamefont
  {Low}(2009)}]{harrow2009random}%
  \BibitemOpen
  \bibfield  {author} {\bibinfo {author} {\bibfnamefont {A.~W.}\ \bibnamefont
  {Harrow}}\ and\ \bibinfo {author} {\bibfnamefont {R.~A.}\ \bibnamefont
  {Low}},\ }\href@noop {} {\bibfield  {journal} {\bibinfo  {journal}
  {Communications in Mathematical Physics}\ }\textbf {\bibinfo {volume}
  {291}},\ \bibinfo {pages} {257} (\bibinfo {year} {2009})}\BibitemShut
  {NoStop}%
\bibitem [{\citenamefont {Brandao}\ \emph
  {et~al.}(2016{\natexlab{a}})\citenamefont {Brandao}, \citenamefont {Harrow},\
  and\ \citenamefont {Horodecki}}]{brandao2016local}%
  \BibitemOpen
  \bibfield  {author} {\bibinfo {author} {\bibfnamefont {F.~G.}\ \bibnamefont
  {Brandao}}, \bibinfo {author} {\bibfnamefont {A.~W.}\ \bibnamefont {Harrow}},
  \ and\ \bibinfo {author} {\bibfnamefont {M.}~\bibnamefont {Horodecki}},\
  }\href@noop {} {\bibfield  {journal} {\bibinfo  {journal} {Communications in
  Mathematical Physics}\ }\textbf {\bibinfo {volume} {346}},\ \bibinfo {pages}
  {397} (\bibinfo {year} {2016}{\natexlab{a}})}\BibitemShut {NoStop}%
\bibitem [{\citenamefont {Hunter-Jones}(2019)}]{stat}%
  \BibitemOpen
  \bibfield  {author} {\bibinfo {author} {\bibfnamefont {N.}~\bibnamefont
  {Hunter-Jones}},\ }\href@noop {} {\  (\bibinfo {year} {2019})},\ \bibinfo
  {note} {preprint at \url{https://arxiv.org/abs/1905.12053}}\BibitemShut
  {NoStop}%
\bibitem [{\citenamefont {Hayden}\ and\ \citenamefont
  {Preskill}(2007)}]{Hayden:2007cs}%
  \BibitemOpen
  \bibfield  {author} {\bibinfo {author} {\bibfnamefont {P.}~\bibnamefont
  {Hayden}}\ and\ \bibinfo {author} {\bibfnamefont {J.}~\bibnamefont
  {Preskill}},\ }\href {\doibase 10.1088/1126-6708/2007/09/120} {\bibfield
  {journal} {\bibinfo  {journal} {JHEP}\ }\textbf {\bibinfo {volume} {09}},\
  \bibinfo {pages} {120} (\bibinfo {year} {2007})}\BibitemShut {NoStop}%
\bibitem [{\citenamefont {Nahum}\ \emph {et~al.}(2018)\citenamefont {Nahum},
  \citenamefont {Vijay},\ and\ \citenamefont {Haah}}]{Nahum:2017yvy}%
  \BibitemOpen
  \bibfield  {author} {\bibinfo {author} {\bibfnamefont {A.}~\bibnamefont
  {Nahum}}, \bibinfo {author} {\bibfnamefont {S.}~\bibnamefont {Vijay}}, \ and\
  \bibinfo {author} {\bibfnamefont {J.}~\bibnamefont {Haah}},\ }\href {\doibase
  10.1103/PhysRevX.8.021014} {\bibfield  {journal} {\bibinfo  {journal} {Phys.
  Rev. X}\ }\textbf {\bibinfo {volume} {8}},\ \bibinfo {pages} {021014}
  (\bibinfo {year} {2018})}\BibitemShut {NoStop}%
\bibitem [{\citenamefont {McClean}\ \emph {et~al.}(2018)\citenamefont
  {McClean}, \citenamefont {Boixo}, \citenamefont {Smelyanskiy}, \citenamefont
  {Babbush},\ and\ \citenamefont {Neven}}]{mcclean2018barren}%
  \BibitemOpen
  \bibfield  {author} {\bibinfo {author} {\bibfnamefont {J.~R.}\ \bibnamefont
  {McClean}}, \bibinfo {author} {\bibfnamefont {S.}~\bibnamefont {Boixo}},
  \bibinfo {author} {\bibfnamefont {V.~N.}\ \bibnamefont {Smelyanskiy}},
  \bibinfo {author} {\bibfnamefont {R.}~\bibnamefont {Babbush}}, \ and\
  \bibinfo {author} {\bibfnamefont {H.}~\bibnamefont {Neven}},\ }\href@noop {}
  {\bibfield  {journal} {\bibinfo  {journal} {Nature communications}\ }\textbf
  {\bibinfo {volume} {9}},\ \bibinfo {pages} {1} (\bibinfo {year}
  {2018})}\BibitemShut {NoStop}%
\bibitem [{\citenamefont {Liu}\ \emph {et~al.}(2021{\natexlab{a}})\citenamefont
  {Liu}, \citenamefont {Tacchino}, \citenamefont {Glick}, \citenamefont
  {Jiang},\ and\ \citenamefont {Mezzacapo}}]{Liu:2021wqr}%
  \BibitemOpen
  \bibfield  {author} {\bibinfo {author} {\bibfnamefont {J.}~\bibnamefont
  {Liu}}, \bibinfo {author} {\bibfnamefont {F.}~\bibnamefont {Tacchino}},
  \bibinfo {author} {\bibfnamefont {J.~R.}\ \bibnamefont {Glick}}, \bibinfo
  {author} {\bibfnamefont {L.}~\bibnamefont {Jiang}}, \ and\ \bibinfo {author}
  {\bibfnamefont {A.}~\bibnamefont {Mezzacapo}},\ }\href@noop {} {\enquote
  {\bibinfo {title} {Representation learning via quantum neural tangent
  kernels},}\ } (\bibinfo {year} {2021}{\natexlab{a}}),\ \bibinfo {note}
  {preprint at \url{https://arxiv.org/abs/2111.04225}}\BibitemShut {NoStop}%
\bibitem [{\citenamefont {Liu}\ \emph {et~al.}(2022{\natexlab{a}})\citenamefont
  {Liu}, \citenamefont {Najafi}, \citenamefont {Sharma}, \citenamefont
  {Tacchino}, \citenamefont {Jiang},\ and\ \citenamefont
  {Mezzacapo}}]{Liu:2022eqa}%
  \BibitemOpen
  \bibfield  {author} {\bibinfo {author} {\bibfnamefont {J.}~\bibnamefont
  {Liu}}, \bibinfo {author} {\bibfnamefont {K.}~\bibnamefont {Najafi}},
  \bibinfo {author} {\bibfnamefont {K.}~\bibnamefont {Sharma}}, \bibinfo
  {author} {\bibfnamefont {F.}~\bibnamefont {Tacchino}}, \bibinfo {author}
  {\bibfnamefont {L.}~\bibnamefont {Jiang}}, \ and\ \bibinfo {author}
  {\bibfnamefont {A.}~\bibnamefont {Mezzacapo}},\ }\href@noop {} {\  (\bibinfo
  {year} {2022}{\natexlab{a}})},\ \bibinfo {note} {preprint at
  \url{https://arxiv.org/abs/2203.16711}}\BibitemShut {NoStop}%
\bibitem [{\citenamefont {Brown}\ and\ \citenamefont
  {Fawzi}(2015)}]{brown2015decoupling}%
  \BibitemOpen
  \bibfield  {author} {\bibinfo {author} {\bibfnamefont {W.}~\bibnamefont
  {Brown}}\ and\ \bibinfo {author} {\bibfnamefont {O.}~\bibnamefont {Fawzi}},\
  }\href@noop {} {\bibfield  {journal} {\bibinfo  {journal} {Communications in
  mathematical physics}\ }\textbf {\bibinfo {volume} {340}},\ \bibinfo {pages}
  {867} (\bibinfo {year} {2015})}\BibitemShut {NoStop}%
\bibitem [{\citenamefont {Liu}(2020)}]{Liu:2020sqb}%
  \BibitemOpen
  \bibfield  {author} {\bibinfo {author} {\bibfnamefont {J.}~\bibnamefont
  {Liu}},\ }\href {\doibase 10.1103/PhysRevResearch.2.043164} {\bibfield
  {journal} {\bibinfo  {journal} {Phys. Rev. Res.}\ }\textbf {\bibinfo {volume}
  {2}},\ \bibinfo {pages} {043164} (\bibinfo {year} {2020})}\BibitemShut
  {NoStop}%
\bibitem [{\citenamefont {Roberts}\ and\ \citenamefont
  {Yoshida}(2017)}]{Roberts:2016hpo}%
  \BibitemOpen
  \bibfield  {author} {\bibinfo {author} {\bibfnamefont {D.~A.}\ \bibnamefont
  {Roberts}}\ and\ \bibinfo {author} {\bibfnamefont {B.}~\bibnamefont
  {Yoshida}},\ }\href {\doibase 10.1007/JHEP04(2017)121} {\bibfield  {journal}
  {\bibinfo  {journal} {JHEP}\ }\textbf {\bibinfo {volume} {04}},\ \bibinfo
  {pages} {121} (\bibinfo {year} {2017})}\BibitemShut {NoStop}%
\bibitem [{\citenamefont {Cotler}\ \emph {et~al.}(2017)\citenamefont {Cotler},
  \citenamefont {Hunter-Jones}, \citenamefont {Liu},\ and\ \citenamefont
  {Yoshida}}]{Cotler:2017jue}%
  \BibitemOpen
  \bibfield  {author} {\bibinfo {author} {\bibfnamefont {J.}~\bibnamefont
  {Cotler}}, \bibinfo {author} {\bibfnamefont {N.}~\bibnamefont
  {Hunter-Jones}}, \bibinfo {author} {\bibfnamefont {J.}~\bibnamefont {Liu}}, \
  and\ \bibinfo {author} {\bibfnamefont {B.}~\bibnamefont {Yoshida}},\ }\href
  {\doibase 10.1007/JHEP11(2017)048} {\bibfield  {journal} {\bibinfo  {journal}
  {JHEP}\ }\textbf {\bibinfo {volume} {11}},\ \bibinfo {pages} {048} (\bibinfo
  {year} {2017})}\BibitemShut {NoStop}%
\bibitem [{\citenamefont {Liu}(2018)}]{Liu:2018hlr}%
  \BibitemOpen
  \bibfield  {author} {\bibinfo {author} {\bibfnamefont {J.}~\bibnamefont
  {Liu}},\ }\href {\doibase 10.1103/PhysRevD.98.086026} {\bibfield  {journal}
  {\bibinfo  {journal} {Phys. Rev. D}\ }\textbf {\bibinfo {volume} {98}},\
  \bibinfo {pages} {086026} (\bibinfo {year} {2018})}\BibitemShut {NoStop}%
\bibitem [{\citenamefont {Brandao}\ and\ \citenamefont
  {Horodecki}(2010)}]{brandao2010exponential}%
  \BibitemOpen
  \bibfield  {author} {\bibinfo {author} {\bibfnamefont {F.~G.}\ \bibnamefont
  {Brandao}}\ and\ \bibinfo {author} {\bibfnamefont {M.}~\bibnamefont
  {Horodecki}},\ }\href@noop {} {\  (\bibinfo {year} {2010})},\ \bibinfo {note}
  {preprint at \url{https://arxiv.org/abs/1010.3654}}\BibitemShut {NoStop}%
\bibitem [{\citenamefont {Harlow}\ and\ \citenamefont
  {Hayden}(2013)}]{harlow2013quantum}%
  \BibitemOpen
  \bibfield  {author} {\bibinfo {author} {\bibfnamefont {D.}~\bibnamefont
  {Harlow}}\ and\ \bibinfo {author} {\bibfnamefont {P.}~\bibnamefont
  {Hayden}},\ }\href@noop {} {\bibfield  {journal} {\bibinfo  {journal}
  {Journal of High Energy Physics}\ }\textbf {\bibinfo {volume} {2013}},\
  \bibinfo {pages} {1} (\bibinfo {year} {2013})}\BibitemShut {NoStop}%
\bibitem [{\citenamefont {Brandao}\ \emph
  {et~al.}(2016{\natexlab{b}})\citenamefont {Brandao}, \citenamefont {Harrow},\
  and\ \citenamefont {Horodecki}}]{brandao2016efficient}%
  \BibitemOpen
  \bibfield  {author} {\bibinfo {author} {\bibfnamefont {F.~G.}\ \bibnamefont
  {Brandao}}, \bibinfo {author} {\bibfnamefont {A.~W.}\ \bibnamefont {Harrow}},
  \ and\ \bibinfo {author} {\bibfnamefont {M.}~\bibnamefont {Horodecki}},\
  }\href@noop {} {\bibfield  {journal} {\bibinfo  {journal} {Physical Review
  Letters}\ }\textbf {\bibinfo {volume} {116}},\ \bibinfo {pages} {170502}
  (\bibinfo {year} {2016}{\natexlab{b}})}\BibitemShut {NoStop}%
\bibitem [{\citenamefont {Ji}\ \emph {et~al.}(2017)\citenamefont {Ji},
  \citenamefont {Liu},\ and\ \citenamefont {Song}}]{ji2017pseudorandom}%
  \BibitemOpen
  \bibfield  {author} {\bibinfo {author} {\bibfnamefont {Z.}~\bibnamefont
  {Ji}}, \bibinfo {author} {\bibfnamefont {Y.-K.}\ \bibnamefont {Liu}}, \ and\
  \bibinfo {author} {\bibfnamefont {F.}~\bibnamefont {Song}},\ }\href@noop {}
  {\  (\bibinfo {year} {2017})},\ \bibinfo {note} {preprint at
  \url{https://arxiv.org/abs/1711.00385}}\BibitemShut {NoStop}%
\bibitem [{\citenamefont {Ananth}\ \emph {et~al.}(2021)\citenamefont {Ananth},
  \citenamefont {Qian},\ and\ \citenamefont {Yuen}}]{ananth2021cryptography}%
  \BibitemOpen
  \bibfield  {author} {\bibinfo {author} {\bibfnamefont {P.}~\bibnamefont
  {Ananth}}, \bibinfo {author} {\bibfnamefont {L.}~\bibnamefont {Qian}}, \ and\
  \bibinfo {author} {\bibfnamefont {H.}~\bibnamefont {Yuen}},\ }\href@noop {}
  {\  (\bibinfo {year} {2021})},\ \bibinfo {note} {preprint at
  \url{https://arxiv.org/abs/2112.10020}}\BibitemShut {NoStop}%
\bibitem [{\citenamefont {{\v{S}}kori{\'c}}(2012)}]{vskoric2012quantum}%
  \BibitemOpen
  \bibfield  {author} {\bibinfo {author} {\bibfnamefont {B.}~\bibnamefont
  {{\v{S}}kori{\'c}}},\ }\href@noop {} {\bibfield  {journal} {\bibinfo
  {journal} {International Journal of Quantum Information}\ }\textbf {\bibinfo
  {volume} {10}},\ \bibinfo {pages} {1250001} (\bibinfo {year}
  {2012})}\BibitemShut {NoStop}%
\bibitem [{\citenamefont {Gianfelici}\ \emph {et~al.}(2020)\citenamefont
  {Gianfelici}, \citenamefont {Kampermann},\ and\ \citenamefont
  {Bru{\ss}}}]{gianfelici2020theoretical}%
  \BibitemOpen
  \bibfield  {author} {\bibinfo {author} {\bibfnamefont {G.}~\bibnamefont
  {Gianfelici}}, \bibinfo {author} {\bibfnamefont {H.}~\bibnamefont
  {Kampermann}}, \ and\ \bibinfo {author} {\bibfnamefont {D.}~\bibnamefont
  {Bru{\ss}}},\ }\href@noop {} {\bibfield  {journal} {\bibinfo  {journal}
  {Physical Review A}\ }\textbf {\bibinfo {volume} {101}},\ \bibinfo {pages}
  {042337} (\bibinfo {year} {2020})}\BibitemShut {NoStop}%
\bibitem [{\citenamefont {Kumar}\ \emph {et~al.}(2021)\citenamefont {Kumar},
  \citenamefont {Mezher},\ and\ \citenamefont {Kashefi}}]{kumar2021efficient}%
  \BibitemOpen
  \bibfield  {author} {\bibinfo {author} {\bibfnamefont {N.}~\bibnamefont
  {Kumar}}, \bibinfo {author} {\bibfnamefont {R.}~\bibnamefont {Mezher}}, \
  and\ \bibinfo {author} {\bibfnamefont {E.}~\bibnamefont {Kashefi}},\
  }\href@noop {} {\  (\bibinfo {year} {2021})},\ \bibinfo {note} {preprint at
  \url{https://arxiv.org/abs/2101.05692}}\BibitemShut {NoStop}%
\bibitem [{\citenamefont {Doosti}\ \emph {et~al.}(2021)\citenamefont {Doosti},
  \citenamefont {Kumar}, \citenamefont {Kashefi},\ and\ \citenamefont
  {Chakraborty}}]{doosti2021connection}%
  \BibitemOpen
  \bibfield  {author} {\bibinfo {author} {\bibfnamefont {M.}~\bibnamefont
  {Doosti}}, \bibinfo {author} {\bibfnamefont {N.}~\bibnamefont {Kumar}},
  \bibinfo {author} {\bibfnamefont {E.}~\bibnamefont {Kashefi}}, \ and\
  \bibinfo {author} {\bibfnamefont {K.}~\bibnamefont {Chakraborty}},\
  }\href@noop {} {\  (\bibinfo {year} {2021})},\ \bibinfo {note} {preprint at
  \url{https://arxiv.org/abs/2110.11724}}\BibitemShut {NoStop}%
\bibitem [{\citenamefont {Arapinis}\ \emph {et~al.}(2021)\citenamefont
  {Arapinis}, \citenamefont {Delavar}, \citenamefont {Doosti},\ and\
  \citenamefont {Kashefi}}]{arapinis2021quantum}%
  \BibitemOpen
  \bibfield  {author} {\bibinfo {author} {\bibfnamefont {M.}~\bibnamefont
  {Arapinis}}, \bibinfo {author} {\bibfnamefont {M.}~\bibnamefont {Delavar}},
  \bibinfo {author} {\bibfnamefont {M.}~\bibnamefont {Doosti}}, \ and\ \bibinfo
  {author} {\bibfnamefont {E.}~\bibnamefont {Kashefi}},\ }\href@noop {}
  {\bibfield  {journal} {\bibinfo  {journal} {Quantum}\ }\textbf {\bibinfo
  {volume} {5}},\ \bibinfo {pages} {475} (\bibinfo {year} {2021})}\BibitemShut
  {NoStop}%
\bibitem [{\citenamefont {Lykov}\ \emph {et~al.}(2021)\citenamefont {Lykov},
  \citenamefont {Chen}, \citenamefont {Chen}, \citenamefont {Keipert},
  \citenamefont {Zhang}, \citenamefont {Gibbs},\ and\ \citenamefont
  {Alexeev}}]{lykov2021performance}%
  \BibitemOpen
  \bibfield  {author} {\bibinfo {author} {\bibfnamefont {D.}~\bibnamefont
  {Lykov}}, \bibinfo {author} {\bibfnamefont {A.}~\bibnamefont {Chen}},
  \bibinfo {author} {\bibfnamefont {H.}~\bibnamefont {Chen}}, \bibinfo {author}
  {\bibfnamefont {K.}~\bibnamefont {Keipert}}, \bibinfo {author} {\bibfnamefont
  {Z.}~\bibnamefont {Zhang}}, \bibinfo {author} {\bibfnamefont
  {T.}~\bibnamefont {Gibbs}}, \ and\ \bibinfo {author} {\bibfnamefont
  {Y.}~\bibnamefont {Alexeev}},\ }in\ \href@noop {} {\emph {\bibinfo
  {booktitle} {2021 IEEE/ACM Second International Workshop on Quantum Computing
  Software (QCS)}}}\ (\bibinfo {organization} {IEEE},\ \bibinfo {year} {2021})\
  pp.\ \bibinfo {pages} {27--34}\BibitemShut {NoStop}%
\bibitem [{\citenamefont {Lykov}\ and\ \citenamefont
  {Alexeev}(2021)}]{lykov_diagonal}%
  \BibitemOpen
  \bibfield  {author} {\bibinfo {author} {\bibfnamefont {D.}~\bibnamefont
  {Lykov}}\ and\ \bibinfo {author} {\bibfnamefont {Y.}~\bibnamefont
  {Alexeev}},\ }\href@noop {} {\enquote {\bibinfo {title} {Importance of
  diagonal gates in tensor network simulations},}\ } (\bibinfo {year} {2021}),\
  \bibinfo {note} {preprint at
  \url{https://arxiv.org/abs/2106.15740}}\BibitemShut {NoStop}%
\bibitem [{\citenamefont {Lykov}\ \emph {et~al.}(2020)\citenamefont {Lykov},
  \citenamefont {Schutski}, \citenamefont {Galda}, \citenamefont {Vinokur},\
  and\ \citenamefont {Alexeev}}]{lykov2021large}%
  \BibitemOpen
  \bibfield  {author} {\bibinfo {author} {\bibfnamefont {D.}~\bibnamefont
  {Lykov}}, \bibinfo {author} {\bibfnamefont {R.}~\bibnamefont {Schutski}},
  \bibinfo {author} {\bibfnamefont {A.}~\bibnamefont {Galda}}, \bibinfo
  {author} {\bibfnamefont {V.}~\bibnamefont {Vinokur}}, \ and\ \bibinfo
  {author} {\bibfnamefont {Y.}~\bibnamefont {Alexeev}},\ }\href@noop {} {\
  (\bibinfo {year} {2020})},\ \bibinfo {note} {preprint at
  \url{https://arxiv.org/abs/2012.02430}}\BibitemShut {NoStop}%
\bibitem [{\citenamefont {Diniz}\ and\ \citenamefont
  {Jonathan}(2011)}]{diniz2011comment}%
  \BibitemOpen
  \bibfield  {author} {\bibinfo {author} {\bibfnamefont {I.~T.}\ \bibnamefont
  {Diniz}}\ and\ \bibinfo {author} {\bibfnamefont {D.}~\bibnamefont
  {Jonathan}},\ }\href@noop {} {\bibfield  {journal} {\bibinfo  {journal}
  {Communications in Mathematical Physics}\ }\textbf {\bibinfo {volume}
  {304}},\ \bibinfo {pages} {281} (\bibinfo {year} {2011})}\BibitemShut
  {NoStop}%
\bibitem [{\citenamefont {Harrow}\ and\ \citenamefont
  {Mehraban}(2018)}]{harrow2018approximate}%
  \BibitemOpen
  \bibfield  {author} {\bibinfo {author} {\bibfnamefont {A.}~\bibnamefont
  {Harrow}}\ and\ \bibinfo {author} {\bibfnamefont {S.}~\bibnamefont
  {Mehraban}},\ }\href@noop {} {\  (\bibinfo {year} {2018})},\ \bibinfo {note}
  {preprint at \url{https://arxiv.org/abs/1809.06957}}\BibitemShut {NoStop}%
\bibitem [{\citenamefont {Nakata}\ \emph {et~al.}(2016)\citenamefont {Nakata},
  \citenamefont {Hirche}, \citenamefont {Koashi},\ and\ \citenamefont
  {Winter}}]{nakata2016efficient}%
  \BibitemOpen
  \bibfield  {author} {\bibinfo {author} {\bibfnamefont {Y.}~\bibnamefont
  {Nakata}}, \bibinfo {author} {\bibfnamefont {C.}~\bibnamefont {Hirche}},
  \bibinfo {author} {\bibfnamefont {M.}~\bibnamefont {Koashi}}, \ and\ \bibinfo
  {author} {\bibfnamefont {A.}~\bibnamefont {Winter}},\ }\href@noop {} {\
  (\bibinfo {year} {2016})},\ \bibinfo {note} {preprint at
  \url{https://arxiv.org/abs/1609.07021}}\BibitemShut {NoStop}%
\bibitem [{\citenamefont {Onorati}\ \emph {et~al.}(2017)\citenamefont
  {Onorati}, \citenamefont {Buerschaper}, \citenamefont {Kliesch},
  \citenamefont {Brown}, \citenamefont {Werner},\ and\ \citenamefont
  {Eisert}}]{onorati2017mixing}%
  \BibitemOpen
  \bibfield  {author} {\bibinfo {author} {\bibfnamefont {E.}~\bibnamefont
  {Onorati}}, \bibinfo {author} {\bibfnamefont {O.}~\bibnamefont
  {Buerschaper}}, \bibinfo {author} {\bibfnamefont {M.}~\bibnamefont
  {Kliesch}}, \bibinfo {author} {\bibfnamefont {W.}~\bibnamefont {Brown}},
  \bibinfo {author} {\bibfnamefont {A.~H.}\ \bibnamefont {Werner}}, \ and\
  \bibinfo {author} {\bibfnamefont {J.}~\bibnamefont {Eisert}},\ }\href@noop {}
  {\bibfield  {journal} {\bibinfo  {journal} {Communications in Mathematical
  Physics}\ }\textbf {\bibinfo {volume} {355}},\ \bibinfo {pages} {905}
  (\bibinfo {year} {2017})}\BibitemShut {NoStop}%
\bibitem [{\citenamefont {Lashkari}\ \emph {et~al.}(2013)\citenamefont
  {Lashkari}, \citenamefont {Stanford}, \citenamefont {Hastings}, \citenamefont
  {Osborne},\ and\ \citenamefont {Hayden}}]{Lashkari:2011yi}%
  \BibitemOpen
  \bibfield  {author} {\bibinfo {author} {\bibfnamefont {N.}~\bibnamefont
  {Lashkari}}, \bibinfo {author} {\bibfnamefont {D.}~\bibnamefont {Stanford}},
  \bibinfo {author} {\bibfnamefont {M.}~\bibnamefont {Hastings}}, \bibinfo
  {author} {\bibfnamefont {T.}~\bibnamefont {Osborne}}, \ and\ \bibinfo
  {author} {\bibfnamefont {P.}~\bibnamefont {Hayden}},\ }\href {\doibase
  10.1007/JHEP04(2013)022} {\bibfield  {journal} {\bibinfo  {journal} {JHEP}\
  }\textbf {\bibinfo {volume} {04}},\ \bibinfo {pages} {022} (\bibinfo {year}
  {2013})}\BibitemShut {NoStop}%
\bibitem [{\citenamefont {Brand\~ao}\ \emph {et~al.}(2021)\citenamefont
  {Brand\~ao}, \citenamefont {Chemissany}, \citenamefont {Hunter-Jones},
  \citenamefont {Kueng},\ and\ \citenamefont {Preskill}}]{Brandao:2019sgy}%
  \BibitemOpen
  \bibfield  {author} {\bibinfo {author} {\bibfnamefont {F.~G. S.~L.}\
  \bibnamefont {Brand\~ao}}, \bibinfo {author} {\bibfnamefont {W.}~\bibnamefont
  {Chemissany}}, \bibinfo {author} {\bibfnamefont {N.}~\bibnamefont
  {Hunter-Jones}}, \bibinfo {author} {\bibfnamefont {R.}~\bibnamefont {Kueng}},
  \ and\ \bibinfo {author} {\bibfnamefont {J.}~\bibnamefont {Preskill}},\
  }\href {\doibase 10.1103/PRXQuantum.2.030316} {\bibfield  {journal} {\bibinfo
   {journal} {PRX Quantum}\ }\textbf {\bibinfo {volume} {2}},\ \bibinfo {pages}
  {030316} (\bibinfo {year} {2021})}\BibitemShut {NoStop}%
\bibitem [{\citenamefont {Brown}\ and\ \citenamefont
  {Susskind}(2018)}]{Brown:2017jil}%
  \BibitemOpen
  \bibfield  {author} {\bibinfo {author} {\bibfnamefont {A.~R.}\ \bibnamefont
  {Brown}}\ and\ \bibinfo {author} {\bibfnamefont {L.}~\bibnamefont
  {Susskind}},\ }\href {\doibase 10.1103/PhysRevD.97.086015} {\bibfield
  {journal} {\bibinfo  {journal} {Phys. Rev. D}\ }\textbf {\bibinfo {volume}
  {97}},\ \bibinfo {pages} {086015} (\bibinfo {year} {2018})}\BibitemShut
  {NoStop}%
\bibitem [{\citenamefont {Susskind}(2018)}]{Susskind:2018fmx}%
  \BibitemOpen
  \bibfield  {author} {\bibinfo {author} {\bibfnamefont {L.}~\bibnamefont
  {Susskind}},\ }\href@noop {} {\  (\bibinfo {year} {2018})},\ \bibinfo {note}
  {preprint at \url{https://arxiv.org/abs/1802.02175}}\BibitemShut {NoStop}%
\bibitem [{\citenamefont {Haferkamp}\ \emph {et~al.}(2022)\citenamefont
  {Haferkamp}, \citenamefont {Faist}, \citenamefont {Kothakonda}, \citenamefont
  {Eisert},\ and\ \citenamefont {Yunger~Halpern}}]{haferkamp2022linear}%
  \BibitemOpen
  \bibfield  {author} {\bibinfo {author} {\bibfnamefont {J.}~\bibnamefont
  {Haferkamp}}, \bibinfo {author} {\bibfnamefont {P.}~\bibnamefont {Faist}},
  \bibinfo {author} {\bibfnamefont {N.~B.}\ \bibnamefont {Kothakonda}},
  \bibinfo {author} {\bibfnamefont {J.}~\bibnamefont {Eisert}}, \ and\ \bibinfo
  {author} {\bibfnamefont {N.}~\bibnamefont {Yunger~Halpern}},\ }\href@noop {}
  {\bibfield  {journal} {\bibinfo  {journal} {Nature Physics}\ ,\ \bibinfo
  {pages} {1}} (\bibinfo {year} {2022})}\BibitemShut {NoStop}%
\bibitem [{\citenamefont {Peruzzo}\ \emph {et~al.}(2014)\citenamefont
  {Peruzzo}, \citenamefont {McClean}, \citenamefont {Shadbolt}, \citenamefont
  {Yung}, \citenamefont {Zhou}, \citenamefont {Love}, \citenamefont
  {Aspuru-Guzik},\ and\ \citenamefont {O?brien}}]{peruzzo2014variational}%
  \BibitemOpen
  \bibfield  {author} {\bibinfo {author} {\bibfnamefont {A.}~\bibnamefont
  {Peruzzo}}, \bibinfo {author} {\bibfnamefont {J.}~\bibnamefont {McClean}},
  \bibinfo {author} {\bibfnamefont {P.}~\bibnamefont {Shadbolt}}, \bibinfo
  {author} {\bibfnamefont {M.-H.}\ \bibnamefont {Yung}}, \bibinfo {author}
  {\bibfnamefont {X.-Q.}\ \bibnamefont {Zhou}}, \bibinfo {author}
  {\bibfnamefont {P.~J.}\ \bibnamefont {Love}}, \bibinfo {author}
  {\bibfnamefont {A.}~\bibnamefont {Aspuru-Guzik}}, \ and\ \bibinfo {author}
  {\bibfnamefont {J.~L.}\ \bibnamefont {O?brien}},\ }\href@noop {} {\bibfield
  {journal} {\bibinfo  {journal} {Nature Communications}\ }\textbf {\bibinfo
  {volume} {5}},\ \bibinfo {pages} {1} (\bibinfo {year} {2014})}\BibitemShut
  {NoStop}%
\bibitem [{\citenamefont {McClean}\ \emph {et~al.}(2016)\citenamefont
  {McClean}, \citenamefont {Romero}, \citenamefont {Babbush},\ and\
  \citenamefont {Aspuru-Guzik}}]{mcclean2016theory}%
  \BibitemOpen
  \bibfield  {author} {\bibinfo {author} {\bibfnamefont {J.~R.}\ \bibnamefont
  {McClean}}, \bibinfo {author} {\bibfnamefont {J.}~\bibnamefont {Romero}},
  \bibinfo {author} {\bibfnamefont {R.}~\bibnamefont {Babbush}}, \ and\
  \bibinfo {author} {\bibfnamefont {A.}~\bibnamefont {Aspuru-Guzik}},\
  }\href@noop {} {\bibfield  {journal} {\bibinfo  {journal} {New Journal of
  Physics}\ }\textbf {\bibinfo {volume} {18}},\ \bibinfo {pages} {023023}
  (\bibinfo {year} {2016})}\BibitemShut {NoStop}%
\bibitem [{\citenamefont {Kandala}\ \emph {et~al.}(2017)\citenamefont
  {Kandala}, \citenamefont {Mezzacapo}, \citenamefont {Temme}, \citenamefont
  {Takita}, \citenamefont {Brink}, \citenamefont {Chow},\ and\ \citenamefont
  {Gambetta}}]{kandala2017hardware}%
  \BibitemOpen
  \bibfield  {author} {\bibinfo {author} {\bibfnamefont {A.}~\bibnamefont
  {Kandala}}, \bibinfo {author} {\bibfnamefont {A.}~\bibnamefont {Mezzacapo}},
  \bibinfo {author} {\bibfnamefont {K.}~\bibnamefont {Temme}}, \bibinfo
  {author} {\bibfnamefont {M.}~\bibnamefont {Takita}}, \bibinfo {author}
  {\bibfnamefont {M.}~\bibnamefont {Brink}}, \bibinfo {author} {\bibfnamefont
  {J.~M.}\ \bibnamefont {Chow}}, \ and\ \bibinfo {author} {\bibfnamefont
  {J.~M.}\ \bibnamefont {Gambetta}},\ }\href@noop {} {\bibfield  {journal}
  {\bibinfo  {journal} {Nature}\ }\textbf {\bibinfo {volume} {549}},\ \bibinfo
  {pages} {242} (\bibinfo {year} {2017})}\BibitemShut {NoStop}%
\bibitem [{\citenamefont {Cerezo}\ \emph {et~al.}(2021)\citenamefont {Cerezo},
  \citenamefont {Arrasmith}, \citenamefont {Babbush}, \citenamefont {Benjamin},
  \citenamefont {Endo}, \citenamefont {Fujii}, \citenamefont {McClean},
  \citenamefont {Mitarai}, \citenamefont {Yuan}, \citenamefont {Cincio} \emph
  {et~al.}}]{cerezo2021variational}%
  \BibitemOpen
  \bibfield  {author} {\bibinfo {author} {\bibfnamefont {M.}~\bibnamefont
  {Cerezo}}, \bibinfo {author} {\bibfnamefont {A.}~\bibnamefont {Arrasmith}},
  \bibinfo {author} {\bibfnamefont {R.}~\bibnamefont {Babbush}}, \bibinfo
  {author} {\bibfnamefont {S.~C.}\ \bibnamefont {Benjamin}}, \bibinfo {author}
  {\bibfnamefont {S.}~\bibnamefont {Endo}}, \bibinfo {author} {\bibfnamefont
  {K.}~\bibnamefont {Fujii}}, \bibinfo {author} {\bibfnamefont {J.~R.}\
  \bibnamefont {McClean}}, \bibinfo {author} {\bibfnamefont {K.}~\bibnamefont
  {Mitarai}}, \bibinfo {author} {\bibfnamefont {X.}~\bibnamefont {Yuan}},
  \bibinfo {author} {\bibfnamefont {L.}~\bibnamefont {Cincio}},  \emph
  {et~al.},\ }\href@noop {} {\bibfield  {journal} {\bibinfo  {journal} {Nature
  Reviews Physics}\ ,\ \bibinfo {pages} {1}} (\bibinfo {year}
  {2021})}\BibitemShut {NoStop}%
\bibitem [{\citenamefont {Farhi}\ \emph {et~al.}(2014)\citenamefont {Farhi},
  \citenamefont {Goldstone},\ and\ \citenamefont {Gutmann}}]{farhi2014quantum}%
  \BibitemOpen
  \bibfield  {author} {\bibinfo {author} {\bibfnamefont {E.}~\bibnamefont
  {Farhi}}, \bibinfo {author} {\bibfnamefont {J.}~\bibnamefont {Goldstone}}, \
  and\ \bibinfo {author} {\bibfnamefont {S.}~\bibnamefont {Gutmann}},\
  }\href@noop {} {\  (\bibinfo {year} {2014})},\ \bibinfo {note} {preprint at
  \url{https://arxiv.org/abs/1411.4028}}\BibitemShut {NoStop}%
\bibitem [{\citenamefont {Wittek}(2014)}]{wittek2014quantum}%
  \BibitemOpen
  \bibfield  {author} {\bibinfo {author} {\bibfnamefont {P.}~\bibnamefont
  {Wittek}},\ }\href@noop {} {\emph {\bibinfo {title} {Quantum machine
  learning: what quantum computing means to data mining}}}\ (\bibinfo
  {publisher} {Academic Press},\ \bibinfo {year} {2014})\BibitemShut {NoStop}%
\bibitem [{\citenamefont {Wiebe}\ \emph {et~al.}(2014)\citenamefont {Wiebe},
  \citenamefont {Kapoor},\ and\ \citenamefont {Svore}}]{wiebe2014quantum}%
  \BibitemOpen
  \bibfield  {author} {\bibinfo {author} {\bibfnamefont {N.}~\bibnamefont
  {Wiebe}}, \bibinfo {author} {\bibfnamefont {A.}~\bibnamefont {Kapoor}}, \
  and\ \bibinfo {author} {\bibfnamefont {K.~M.}\ \bibnamefont {Svore}},\
  }\href@noop {} {\enquote {\bibinfo {title} {Quantum deep learning},}\ }
  (\bibinfo {year} {2014}),\ \bibinfo {note} {preprint at
  \url{https://arxiv.org/abs/1412.3489}}\BibitemShut {NoStop}%
\bibitem [{\citenamefont {Biamonte}\ \emph {et~al.}(2017)\citenamefont
  {Biamonte}, \citenamefont {Wittek}, \citenamefont {Pancotti}, \citenamefont
  {Rebentrost}, \citenamefont {Wiebe},\ and\ \citenamefont
  {Lloyd}}]{biamonte2017quantum}%
  \BibitemOpen
  \bibfield  {author} {\bibinfo {author} {\bibfnamefont {J.}~\bibnamefont
  {Biamonte}}, \bibinfo {author} {\bibfnamefont {P.}~\bibnamefont {Wittek}},
  \bibinfo {author} {\bibfnamefont {N.}~\bibnamefont {Pancotti}}, \bibinfo
  {author} {\bibfnamefont {P.}~\bibnamefont {Rebentrost}}, \bibinfo {author}
  {\bibfnamefont {N.}~\bibnamefont {Wiebe}}, \ and\ \bibinfo {author}
  {\bibfnamefont {S.}~\bibnamefont {Lloyd}},\ }\href@noop {} {\bibfield
  {journal} {\bibinfo  {journal} {Nature}\ }\textbf {\bibinfo {volume} {549}},\
  \bibinfo {pages} {195} (\bibinfo {year} {2017})}\BibitemShut {NoStop}%
\bibitem [{\citenamefont {Schuld}\ and\ \citenamefont
  {Killoran}(2019)}]{schuld2019quantum}%
  \BibitemOpen
  \bibfield  {author} {\bibinfo {author} {\bibfnamefont {M.}~\bibnamefont
  {Schuld}}\ and\ \bibinfo {author} {\bibfnamefont {N.}~\bibnamefont
  {Killoran}},\ }\href@noop {} {\bibfield  {journal} {\bibinfo  {journal}
  {Physical Review Letters}\ }\textbf {\bibinfo {volume} {122}},\ \bibinfo
  {pages} {040504} (\bibinfo {year} {2019})}\BibitemShut {NoStop}%
\bibitem [{\citenamefont {Havl{\'\i}{\v{c}}ek}\ \emph
  {et~al.}(2019)\citenamefont {Havl{\'\i}{\v{c}}ek}, \citenamefont
  {C{\'o}rcoles}, \citenamefont {Temme}, \citenamefont {Harrow}, \citenamefont
  {Kandala}, \citenamefont {Chow},\ and\ \citenamefont
  {Gambetta}}]{havlivcek2019supervised}%
  \BibitemOpen
  \bibfield  {author} {\bibinfo {author} {\bibfnamefont {V.}~\bibnamefont
  {Havl{\'\i}{\v{c}}ek}}, \bibinfo {author} {\bibfnamefont {A.~D.}\
  \bibnamefont {C{\'o}rcoles}}, \bibinfo {author} {\bibfnamefont
  {K.}~\bibnamefont {Temme}}, \bibinfo {author} {\bibfnamefont {A.~W.}\
  \bibnamefont {Harrow}}, \bibinfo {author} {\bibfnamefont {A.}~\bibnamefont
  {Kandala}}, \bibinfo {author} {\bibfnamefont {J.~M.}\ \bibnamefont {Chow}}, \
  and\ \bibinfo {author} {\bibfnamefont {J.~M.}\ \bibnamefont {Gambetta}},\
  }\href@noop {} {\bibfield  {journal} {\bibinfo  {journal} {Nature}\ }\textbf
  {\bibinfo {volume} {567}},\ \bibinfo {pages} {209} (\bibinfo {year}
  {2019})}\BibitemShut {NoStop}%
\bibitem [{\citenamefont {Liu}\ \emph {et~al.}(2021{\natexlab{b}})\citenamefont
  {Liu}, \citenamefont {Arunachalam},\ and\ \citenamefont
  {Temme}}]{liu2021rigorous}%
  \BibitemOpen
  \bibfield  {author} {\bibinfo {author} {\bibfnamefont {Y.}~\bibnamefont
  {Liu}}, \bibinfo {author} {\bibfnamefont {S.}~\bibnamefont {Arunachalam}}, \
  and\ \bibinfo {author} {\bibfnamefont {K.}~\bibnamefont {Temme}},\
  }\href@noop {} {\bibfield  {journal} {\bibinfo  {journal} {Nature Physics}\
  ,\ \bibinfo {pages} {1}} (\bibinfo {year} {2021}{\natexlab{b}})}\BibitemShut
  {NoStop}%
\bibitem [{\citenamefont {Liu}(2021)}]{Liu:2021ohs}%
  \BibitemOpen
  \bibfield  {author} {\bibinfo {author} {\bibfnamefont {J.}~\bibnamefont
  {Liu}},\ }\emph {\bibinfo {title} {{Does [Richard Feynman] Dream of Electric
  Sheep? Topics on Quantum Field Theory, Quantum Computing, and Computer
  Science}}},\ \href {\doibase 10.7907/adtc-ss13} {Ph.D. thesis},\ \bibinfo
  {school} {Caltech} (\bibinfo {year} {2021})\BibitemShut {NoStop}%
\bibitem [{\citenamefont {Farhi}\ and\ \citenamefont
  {Neven}(2018)}]{farhi2018classification}%
  \BibitemOpen
  \bibfield  {author} {\bibinfo {author} {\bibfnamefont {E.}~\bibnamefont
  {Farhi}}\ and\ \bibinfo {author} {\bibfnamefont {H.}~\bibnamefont {Neven}},\
  }\href@noop {} {\  (\bibinfo {year} {2018})},\ \bibinfo {note} {preprint at
  \url{https://arxiv.org/abs/1802.06002}}\BibitemShut {NoStop}%
\bibitem [{\citenamefont {Gross}\ \emph {et~al.}(2007)\citenamefont {Gross},
  \citenamefont {Audenaert},\ and\ \citenamefont {Eisert}}]{frame_potential}%
  \BibitemOpen
  \bibfield  {author} {\bibinfo {author} {\bibfnamefont {D.}~\bibnamefont
  {Gross}}, \bibinfo {author} {\bibfnamefont {K.}~\bibnamefont {Audenaert}}, \
  and\ \bibinfo {author} {\bibfnamefont {J.}~\bibnamefont {Eisert}},\ }\href
  {\doibase 10.1063/1.2716992} {\bibfield  {journal} {\bibinfo  {journal}
  {Journal of Mathematical Physics}\ }\textbf {\bibinfo {volume} {48}},\
  \bibinfo {pages} {052104} (\bibinfo {year} {2007})}\BibitemShut {NoStop}%
\bibitem [{\citenamefont {Schutski}\ \emph {et~al.}(2020)\citenamefont
  {Schutski}, \citenamefont {Lykov},\ and\ \citenamefont
  {Oseledets}}]{Schutski}%
  \BibitemOpen
  \bibfield  {author} {\bibinfo {author} {\bibfnamefont {R.}~\bibnamefont
  {Schutski}}, \bibinfo {author} {\bibfnamefont {D.}~\bibnamefont {Lykov}}, \
  and\ \bibinfo {author} {\bibfnamefont {I.}~\bibnamefont {Oseledets}},\ }\href
  {\doibase 10.1103/PhysRevA.101.042335} {\bibfield  {journal} {\bibinfo
  {journal} {Phys. Rev. A}\ }\textbf {\bibinfo {volume} {101}},\ \bibinfo
  {pages} {042335} (\bibinfo {year} {2020})}\BibitemShut {NoStop}%
\bibitem [{\citenamefont {Wang}\ \emph {et~al.}(2020)\citenamefont {Wang},
  \citenamefont {Hu}, \citenamefont {Sanders},\ and\ \citenamefont
  {Kais}}]{qudits}%
  \BibitemOpen
  \bibfield  {author} {\bibinfo {author} {\bibfnamefont {Y.}~\bibnamefont
  {Wang}}, \bibinfo {author} {\bibfnamefont {Z.}~\bibnamefont {Hu}}, \bibinfo
  {author} {\bibfnamefont {B.~C.}\ \bibnamefont {Sanders}}, \ and\ \bibinfo
  {author} {\bibfnamefont {S.}~\bibnamefont {Kais}},\ }\href {\doibase
  10.3389/fphy.2020.589504} {\bibfield  {journal} {\bibinfo  {journal} {Front.
  Phys.}\ }\textbf {\bibinfo {volume} {8}} (\bibinfo {year} {2020}),\
  10.3389/fphy.2020.589504}\BibitemShut {NoStop}%
\bibitem [{\citenamefont {Grossi}\ \emph {et~al.}(2022)\citenamefont {Grossi},
  \citenamefont {Kiss}, \citenamefont {De~Luca}, \citenamefont {Zollo},
  \citenamefont {Gremese},\ and\ \citenamefont {Mandarino}}]{grossi2022finite}%
  \BibitemOpen
  \bibfield  {author} {\bibinfo {author} {\bibfnamefont {M.}~\bibnamefont
  {Grossi}}, \bibinfo {author} {\bibfnamefont {O.}~\bibnamefont {Kiss}},
  \bibinfo {author} {\bibfnamefont {F.}~\bibnamefont {De~Luca}}, \bibinfo
  {author} {\bibfnamefont {C.}~\bibnamefont {Zollo}}, \bibinfo {author}
  {\bibfnamefont {I.}~\bibnamefont {Gremese}}, \ and\ \bibinfo {author}
  {\bibfnamefont {A.}~\bibnamefont {Mandarino}},\ }\href@noop {} {\  (\bibinfo
  {year} {2022})},\ \bibinfo {note} {preprint at
  \url{https://arxiv.org/abs/2208.02731}}\BibitemShut {NoStop}%
\bibitem [{\citenamefont {Nakaji}\ and\ \citenamefont
  {Yamamoto}(2021)}]{nakaji2021expressibility}%
  \BibitemOpen
  \bibfield  {author} {\bibinfo {author} {\bibfnamefont {K.}~\bibnamefont
  {Nakaji}}\ and\ \bibinfo {author} {\bibfnamefont {N.}~\bibnamefont
  {Yamamoto}},\ }\href@noop {} {\bibfield  {journal} {\bibinfo  {journal}
  {Quantum}\ }\textbf {\bibinfo {volume} {5}},\ \bibinfo {pages} {434}
  (\bibinfo {year} {2021})}\BibitemShut {NoStop}%
\bibitem [{\citenamefont {Du}\ \emph {et~al.}(2022)\citenamefont {Du},
  \citenamefont {Huang}, \citenamefont {You}, \citenamefont {Hsieh},\ and\
  \citenamefont {Tao}}]{Du2022search}%
  \BibitemOpen
  \bibfield  {author} {\bibinfo {author} {\bibfnamefont {Y.}~\bibnamefont
  {Du}}, \bibinfo {author} {\bibfnamefont {T.}~\bibnamefont {Huang}}, \bibinfo
  {author} {\bibfnamefont {S.}~\bibnamefont {You}}, \bibinfo {author}
  {\bibfnamefont {M.-H.}\ \bibnamefont {Hsieh}}, \ and\ \bibinfo {author}
  {\bibfnamefont {D.}~\bibnamefont {Tao}},\ }\href@noop {} {\bibfield
  {journal} {\bibinfo  {journal} {npj Quantum Inf}\ }\textbf {\bibinfo {volume}
  {8}},\ \bibinfo {pages} {62} (\bibinfo {year} {2022})}\BibitemShut {NoStop}%
\bibitem [{\citenamefont {Sim}\ \emph {et~al.}(2019)\citenamefont {Sim},
  \citenamefont {Johnson},\ and\ \citenamefont
  {Aspuru-Guzik}}]{expressibility}%
  \BibitemOpen
  \bibfield  {author} {\bibinfo {author} {\bibfnamefont {S.}~\bibnamefont
  {Sim}}, \bibinfo {author} {\bibfnamefont {P.~D.}\ \bibnamefont {Johnson}}, \
  and\ \bibinfo {author} {\bibfnamefont {A.}~\bibnamefont {Aspuru-Guzik}},\
  }\href {\doibase https://doi.org/10.1002/qute.201900070} {\bibfield
  {journal} {\bibinfo  {journal} {Advanced Quantum Technologies}\ }\textbf
  {\bibinfo {volume} {2}},\ \bibinfo {pages} {1900070} (\bibinfo {year}
  {2019})}\BibitemShut {NoStop}%
\bibitem [{\citenamefont {Liu}\ \emph {et~al.}(2022{\natexlab{b}})\citenamefont
  {Liu}, \citenamefont {Angone}, \citenamefont {Shaydulin}, \citenamefont
  {Safro}, \citenamefont {Alexeev},\ and\ \citenamefont
  {Cincio}}]{liu2022layer}%
  \BibitemOpen
  \bibfield  {author} {\bibinfo {author} {\bibfnamefont {X.}~\bibnamefont
  {Liu}}, \bibinfo {author} {\bibfnamefont {A.}~\bibnamefont {Angone}},
  \bibinfo {author} {\bibfnamefont {R.}~\bibnamefont {Shaydulin}}, \bibinfo
  {author} {\bibfnamefont {I.}~\bibnamefont {Safro}}, \bibinfo {author}
  {\bibfnamefont {Y.}~\bibnamefont {Alexeev}}, \ and\ \bibinfo {author}
  {\bibfnamefont {L.}~\bibnamefont {Cincio}},\ }\href@noop {} {\bibfield
  {journal} {\bibinfo  {journal} {IEEE Transactions on Quantum Engineering}\
  }\textbf {\bibinfo {volume} {3}},\ \bibinfo {pages} {1} (\bibinfo {year}
  {2022}{\natexlab{b}})}\BibitemShut {NoStop}%
\bibitem [{\citenamefont {Holmes}\ \emph {et~al.}(2022)\citenamefont {Holmes},
  \citenamefont {Sharma}, \citenamefont {Cerezo},\ and\ \citenamefont
  {Coles}}]{expressibility_barren}%
  \BibitemOpen
  \bibfield  {author} {\bibinfo {author} {\bibfnamefont {Z.}~\bibnamefont
  {Holmes}}, \bibinfo {author} {\bibfnamefont {K.}~\bibnamefont {Sharma}},
  \bibinfo {author} {\bibfnamefont {M.}~\bibnamefont {Cerezo}}, \ and\ \bibinfo
  {author} {\bibfnamefont {P.~J.}\ \bibnamefont {Coles}},\ }\href {\doibase
  10.1103/PRXQuantum.3.010313} {\bibfield  {journal} {\bibinfo  {journal} {PRX
  Quantum}\ }\textbf {\bibinfo {volume} {3}},\ \bibinfo {pages} {010313}
  (\bibinfo {year} {2022})}\BibitemShut {NoStop}%
\bibitem [{\citenamefont {Datta}\ \emph {et~al.}(2005)\citenamefont {Datta},
  \citenamefont {Flammia},\ and\ \citenamefont {Caves}}]{dqc1}%
  \BibitemOpen
  \bibfield  {author} {\bibinfo {author} {\bibfnamefont {A.}~\bibnamefont
  {Datta}}, \bibinfo {author} {\bibfnamefont {S.~T.}\ \bibnamefont {Flammia}},
  \ and\ \bibinfo {author} {\bibfnamefont {C.~M.}\ \bibnamefont {Caves}},\
  }\href@noop {} {\bibfield  {journal} {\bibinfo  {journal} {Physical Review
  A}\ }\textbf {\bibinfo {volume} {72}},\ \bibinfo {pages} {042316} (\bibinfo
  {year} {2005})}\BibitemShut {NoStop}%
\bibitem [{\citenamefont {Yuan}\ \emph {et~al.}(2021)\citenamefont {Yuan},
  \citenamefont {Sun}, \citenamefont {Liu}, \citenamefont {Zhao},\ and\
  \citenamefont {Zhou}}]{Yuan:2020xmq}%
  \BibitemOpen
  \bibfield  {author} {\bibinfo {author} {\bibfnamefont {X.}~\bibnamefont
  {Yuan}}, \bibinfo {author} {\bibfnamefont {J.}~\bibnamefont {Sun}}, \bibinfo
  {author} {\bibfnamefont {J.}~\bibnamefont {Liu}}, \bibinfo {author}
  {\bibfnamefont {Q.}~\bibnamefont {Zhao}}, \ and\ \bibinfo {author}
  {\bibfnamefont {Y.}~\bibnamefont {Zhou}},\ }\href {\doibase
  10.1103/PhysRevLett.127.040501} {\bibfield  {journal} {\bibinfo  {journal}
  {Phys. Rev. Lett.}\ }\textbf {\bibinfo {volume} {127}},\ \bibinfo {pages}
  {040501} (\bibinfo {year} {2021})}\BibitemShut {NoStop}%
\bibitem [{\citenamefont {Milsted}\ \emph {et~al.}(2022)\citenamefont
  {Milsted}, \citenamefont {Liu}, \citenamefont {Preskill},\ and\ \citenamefont
  {Vidal}}]{Milsted:2020jmf}%
  \BibitemOpen
  \bibfield  {author} {\bibinfo {author} {\bibfnamefont {A.}~\bibnamefont
  {Milsted}}, \bibinfo {author} {\bibfnamefont {J.}~\bibnamefont {Liu}},
  \bibinfo {author} {\bibfnamefont {J.}~\bibnamefont {Preskill}}, \ and\
  \bibinfo {author} {\bibfnamefont {G.}~\bibnamefont {Vidal}},\ }\href
  {\doibase 10.1103/PRXQuantum.3.020316} {\bibfield  {journal} {\bibinfo
  {journal} {PRX Quantum}\ }\textbf {\bibinfo {volume} {3}},\ \bibinfo {pages}
  {020316} (\bibinfo {year} {2022})}\BibitemShut {NoStop}%
\bibitem [{\citenamefont {Liu}\ \emph {et~al.}(2022{\natexlab{c}})\citenamefont
  {Liu}, \citenamefont {Liu}, \citenamefont {Liu}, \citenamefont {Makhanov},
  \citenamefont {Lykov}, \citenamefont {Apte},\ and\ \citenamefont
  {Alexeev}}]{liu2022embedding}%
  \BibitemOpen
  \bibfield  {author} {\bibinfo {author} {\bibfnamefont {H.}~\bibnamefont
  {Liu}}, \bibinfo {author} {\bibfnamefont {J.}~\bibnamefont {Liu}}, \bibinfo
  {author} {\bibfnamefont {R.}~\bibnamefont {Liu}}, \bibinfo {author}
  {\bibfnamefont {H.}~\bibnamefont {Makhanov}}, \bibinfo {author}
  {\bibfnamefont {D.}~\bibnamefont {Lykov}}, \bibinfo {author} {\bibfnamefont
  {A.}~\bibnamefont {Apte}}, \ and\ \bibinfo {author} {\bibfnamefont
  {Y.}~\bibnamefont {Alexeev}},\ }\href@noop {} {\  (\bibinfo {year}
  {2022}{\natexlab{c}})},\ \bibinfo {note} {preprint at
  \url{https://arxiv.org/abs/2204.04550}}\BibitemShut {NoStop}%
\bibitem [{\citenamefont {Jing}\ \emph {et~al.}(2017)\citenamefont {Jing},
  \citenamefont {Shen}, \citenamefont {Dubcek}, \citenamefont {Peurifoy},
  \citenamefont {Skirlo}, \citenamefont {LeCun}, \citenamefont {Tegmark},\ and\
  \citenamefont {Solja{\v{c}}i{\'c}}}]{jing2017tunable}%
  \BibitemOpen
  \bibfield  {author} {\bibinfo {author} {\bibfnamefont {L.}~\bibnamefont
  {Jing}}, \bibinfo {author} {\bibfnamefont {Y.}~\bibnamefont {Shen}}, \bibinfo
  {author} {\bibfnamefont {T.}~\bibnamefont {Dubcek}}, \bibinfo {author}
  {\bibfnamefont {J.}~\bibnamefont {Peurifoy}}, \bibinfo {author}
  {\bibfnamefont {S.}~\bibnamefont {Skirlo}}, \bibinfo {author} {\bibfnamefont
  {Y.}~\bibnamefont {LeCun}}, \bibinfo {author} {\bibfnamefont
  {M.}~\bibnamefont {Tegmark}}, \ and\ \bibinfo {author} {\bibfnamefont
  {M.}~\bibnamefont {Solja{\v{c}}i{\'c}}},\ }in\ \href@noop {} {\emph {\bibinfo
  {booktitle} {International Conference on Machine Learning}}}\ (\bibinfo
  {organization} {PMLR},\ \bibinfo {year} {2017})\ pp.\ \bibinfo {pages}
  {1733--1741}\BibitemShut {NoStop}%
\bibitem [{\citenamefont {Laporte}(2020)}]{eunn}%
  \BibitemOpen
  \bibfield  {author} {\bibinfo {author} {\bibfnamefont {F.}~\bibnamefont
  {Laporte}},\ }\href@noop {} {\enquote {\bibinfo {title} {torch\_eunn},}\ }
  (\bibinfo {year} {2020}),\ \bibinfo {note}
  {\url{https://github.com/flaport/torch\_eunn}}\BibitemShut {NoStop}%
\bibitem [{\citenamefont {Reck}\ \emph {et~al.}(1994)\citenamefont {Reck},
  \citenamefont {Zeilinger}, \citenamefont {Bernstein},\ and\ \citenamefont
  {Bertani}}]{Reck1994}%
  \BibitemOpen
  \bibfield  {author} {\bibinfo {author} {\bibfnamefont {M.}~\bibnamefont
  {Reck}}, \bibinfo {author} {\bibfnamefont {A.}~\bibnamefont {Zeilinger}},
  \bibinfo {author} {\bibfnamefont {H.~J.}\ \bibnamefont {Bernstein}}, \ and\
  \bibinfo {author} {\bibfnamefont {P.}~\bibnamefont {Bertani}},\ }\href
  {\doibase 10.1103/PhysRevLett.73.58} {\bibfield  {journal} {\bibinfo
  {journal} {Phys. Rev. Lett.}\ }\textbf {\bibinfo {volume} {73}},\ \bibinfo
  {pages} {58} (\bibinfo {year} {1994})}\BibitemShut {NoStop}%
\bibitem [{\citenamefont {Clements}\ \emph {et~al.}(2016)\citenamefont
  {Clements}, \citenamefont {Humphreys}, \citenamefont {Metcalf}, \citenamefont
  {Kolthammer},\ and\ \citenamefont {Walmsley}}]{Clements2016}%
  \BibitemOpen
  \bibfield  {author} {\bibinfo {author} {\bibfnamefont {W.~R.}\ \bibnamefont
  {Clements}}, \bibinfo {author} {\bibfnamefont {P.~C.}\ \bibnamefont
  {Humphreys}}, \bibinfo {author} {\bibfnamefont {B.~J.}\ \bibnamefont
  {Metcalf}}, \bibinfo {author} {\bibfnamefont {W.~S.}\ \bibnamefont
  {Kolthammer}}, \ and\ \bibinfo {author} {\bibfnamefont {I.~A.}\ \bibnamefont
  {Walmsley}},\ }\href {\doibase 10.1364/OPTICA.3.001460} {\bibfield  {journal}
  {\bibinfo  {journal} {Optica}\ }\textbf {\bibinfo {volume} {3}},\ \bibinfo
  {pages} {1460} (\bibinfo {year} {2016})}\BibitemShut {NoStop}%
\bibitem [{\citenamefont {Liu}(2022{\natexlab{a}})}]{fp}%
  \BibitemOpen
  \bibfield  {author} {\bibinfo {author} {\bibfnamefont {M.}~\bibnamefont
  {Liu}},\ }\href@noop {} {\enquote {\bibinfo {title} {Frame\_potential},}\ }
  (\bibinfo {year} {2022}{\natexlab{a}}),\ \bibinfo {note}
  {\url{https://github.com/sss441803/Frame_Potential}}\BibitemShut {NoStop}%
\bibitem [{\citenamefont {Liu}(2022{\natexlab{b}})}]{qtensor_ai}%
  \BibitemOpen
  \bibfield  {author} {\bibinfo {author} {\bibfnamefont {M.}~\bibnamefont
  {Liu}},\ }\href@noop {} {\enquote {\bibinfo {title} {Qtensorai},}\ }
  (\bibinfo {year} {2022}{\natexlab{b}}),\ \bibinfo {note}
  {\url{https://github.com/sss441803/QTensorAI}}\BibitemShut {NoStop}%
\bibitem [{\citenamefont {Lykov}(2020)}]{qtensor}%
  \BibitemOpen
  \bibfield  {author} {\bibinfo {author} {\bibfnamefont {D.}~\bibnamefont
  {Lykov}},\ }\href@noop {} {\enquote {\bibinfo {title} {Qtensor},}\ }
  (\bibinfo {year} {2020}),\ \bibinfo {note}
  {\url{https://github.com/danlkv/QTensor}}\BibitemShut {NoStop}%
\end{thebibliography}%


%

\section*{Acknowledgements}
This material is based upon work supported by the U.S. Department of Energy, Office of Science, National Quantum Information Science Research Centers. This work was completed in part with resources provided by the University of Chicago Research Computing Center. We thank Jens Eisert and Danylo Lykov for useful discussions.

Y.A. acknowledges support from the U.S. Department of Energy, Office of Science, under contract DE-AC02-06CH11357 at Argonne National Laboratory. \textcolor{black}{This research was developed with funding from the Defense Advanced Research Projects Agency (DARPA). The views, opinions and/or findings expressed are those of the author and should not be interpreted as representing the official views or policies of the Department of Defense or the U.S. Government.}

J.L. is supported in part by International Business Machines (IBM) Quantum through the Chicago Quantum Exchange and by the Pritzker School of Molecular Engineering at the University of Chicago through AFOSR MURI (FA9550-21-1-0209). J.L. is also serving as a scientific advisor in qBraid Co. 

L.J. acknowledges support from ARO (W911NF-18-1-0020, W911NF-18-1-0212), ARO MURI (W911NF-16-1-0349, W911NF-21-1-0325), AFOSR MURI (FA9550-19-1-0399, FA9550-21-1-0209), AFRL (FA8649-21-P-0781), DoE Q-NEXT, NSF (OMA-1936118, EEC-1941583, OMA-2137642), NTT Research, and the Packard Foundation (2020-71479).

\section{Author information}
\subsection{Contributions}
J.L. conceived the idea and wrote the majority of the introduction and discussion. M.L developed and performed all numerical simulations and wrote the majority of the introduction to frame potential, results and methods. Y.A. and L.J contributed novel ideas and provided numerous scientific and writing improvements to the paper. M.L., J.L., Y.A., and L.J. all participated in discussions that shaped the project in a substantial manner and the understanding of its broader impact.

\section{Ethics declarations}
The authors declare no competing financial or non-financial interests.


\end{document}


\title[Random Quantum Circuits]{Supplementary Materials of Estimating the randomness of quantum circuit ensembles up to 50 qubits}


\author*[1,2]{\fnm{Minzhao} \sur{Liu}}\email{mliu6@uchicago.edu}

\author*[3,4,5,6]{\fnm{Junyu} \sur{Liu}}\email{junyuliu@uchicago.edu}

\author[2,4]{\fnm{Yuri} \sur{Alexeev}}\email{yuri@anl.gov}

\author[3,4]{\fnm{Liang} \sur{Jiang}}\email{liangjiang@uchicago.edu}

\affil[1]{\orgdiv{Department of Physics}, \orgname{University of Chicago}, \orgaddress{\city{Chicago}, \state{IL} \postcode{60637}, \country{USA}}}

\affil[2]{\orgdiv{Computational Science Division}, \orgname{Argonne National Laboratory}, \orgaddress{\city{Lemont}, \state{IL} \postcode{60439}, \country{USA}}}

\affil[3]{\orgdiv{Pritzker School of Molecular Engineering}, \orgname{University of Chicago}, \orgaddress{\city{Chicago}, \state{IL} \postcode{60637}, \country{USA}}}

\affil[4]{\orgname{Chicago Quantum Exchange}, \orgaddress{\city{Chicago}, \state{IL} \postcode{60637}, \country{USA}}}

\affil[5]{\orgdiv{Kadanoff Center for Theoretical Physics}, \orgname{University of Chicago}, \orgaddress{\city{Chicago}, \state{IL} \postcode{60637}, \country{USA}}}

\affil[6]{\orgname{qBraid Co.}, \orgaddress{\street{Harper Court 5235}, \city{Chicago}, \state{IL} \postcode{60615}, \country{USA}}}

\maketitle


\newpage

\section{Supplementary Methods}

\subsection{Determination of sample sizes}
Since the simulation of large circuits is expensive, we adaptively terminate sampling simulations based on the estimated standard error in the frame potential according to various rules we implement. For example, the simulation stops if the frame potential is larger than the Haar value by a few standard errors or takes too long. If the estimated frame potential is less than the Haar value or smaller than that of a lower depth circuit, the simulation would run longer to steer away from unphysical results.

\subsection{Bootstrapping for uncertainty quantification}

As discussed in the main manuscript, we use bootstrapping to analyze the uncertainties in the data. This is because of the highly asymmetric nature of the error due the highly skewed distribution of $\vert\text{Tr}\left(U^{\dagger}V\right)\vert^k$ (rare cases where $U$ and $V$ collide, making the quantity is very large, and the skewedness worsens for larger $k$s). Performing naive error propagation using the standard error without taking asymmetry into account after curve fitting and solving for intersection is therefore unreliable.

Bootstrapping is a good method for uncertainty quantification because it does not assume the distribution of the estimator. Given a set $S$ of $N$ samples obtained from population $P$, we are interested in the estimator value $E$ and its uncertainty. Bootstrapping resamples $N$ samples with replacement from $S$ to form $S_i$, and calculates $E_i$ multiple times to form a set of bootstrap samples $\{E_i\}$. This means each $S_i$ must repeat and omit some samples from $S$. Assuming $S$ is a good representation of the population $P$, it is as if each $E_i$ is sampled from $P$. Therefore, the distribution of $E_i$ should approximate the actual distribution of $E$.

In our analysis, $P$ is the trace distributed over the ans{\"a}tz measure, $E$ can be the frame potential, layers needed to reach $\epsilon$, etc. We use 300 bootstrap samples in our analysis.

\subsection{Choice of curve fitting region}

Assuming exponential scaling of $\mathcal{F}$ in $l$, we can fit an exponential curve. As can be seen from Supplementary Figure 1, 4, 6, and 8 of the main manuscript, $\log{\mathcal{F}}$ does not scale linearly until large $l$ or small $\mathcal{F}$. Therefore, we choose layers that has $\frac{\mathcal{F}_\mathcal{E}^{(k)}-\mathcal{F}_{\text{Haar}}^{(k)}}{\mathcal{F}_{\text{Haar}}^{(k)}}<5$, which by eye roughly corresponds to the exponential scaling regimes as shown in Supplementary Figure 4, 6, and 8 of the main manuscript.

\section{Supplementary Figures}

\subsection{Sample sizes}
We show in Supplementary Figure \ref{circuit samples} the number of independently sampled circuits used to obtain the frame potential for a single data point. We compute and store the individual trace values $\text{Tr}(U^{\dagger}V)$ for each sampled circuit, and use the same trace values to compute frame potentials with different $k$ values. Therefore, the number of samples has only qubit and layer dependence. As we can see, most data points are obtained with at least 1000 samples.

\begin{figure} [ht]
   \begin{center}
   \includegraphics[width=8cm]{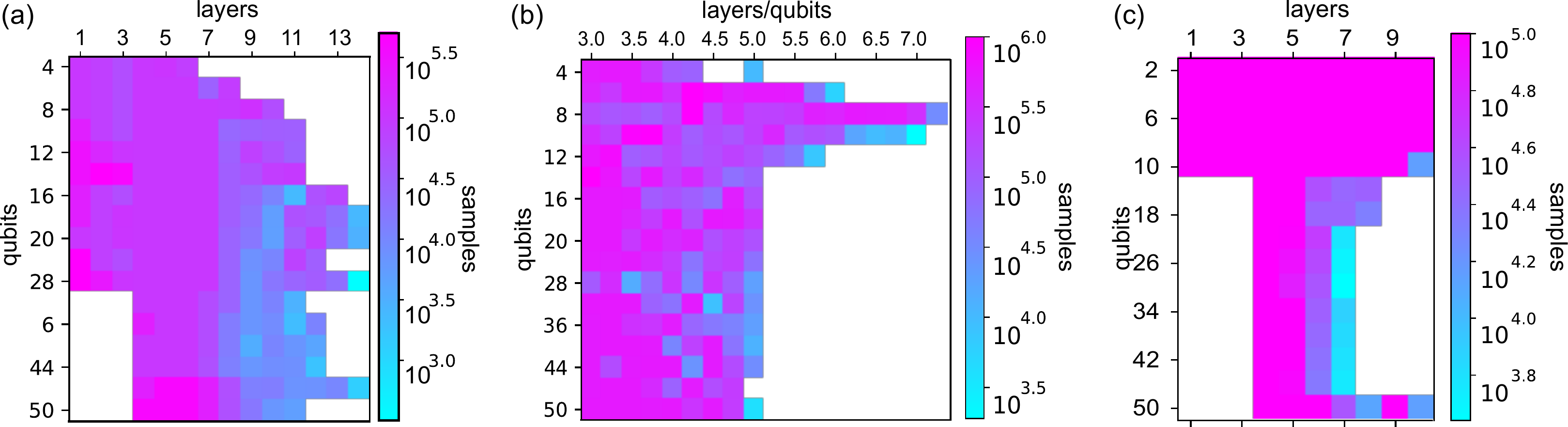}
   \end{center}
   \caption{Number of circuit samples used for estimating the frame potential for each circuit configuration. (a) Parallel random unitary. (b) Local random unitary. (c) The Hardware-efficient ans{\"a}tz with CNOT gates.}
   { \label{circuit samples}
}
   \end{figure}
   
\subsection{Detailed uncertainties obtained from bootstrapping}

\begin{figure} [ht]
   \begin{center}
   \includegraphics[width=7cm]{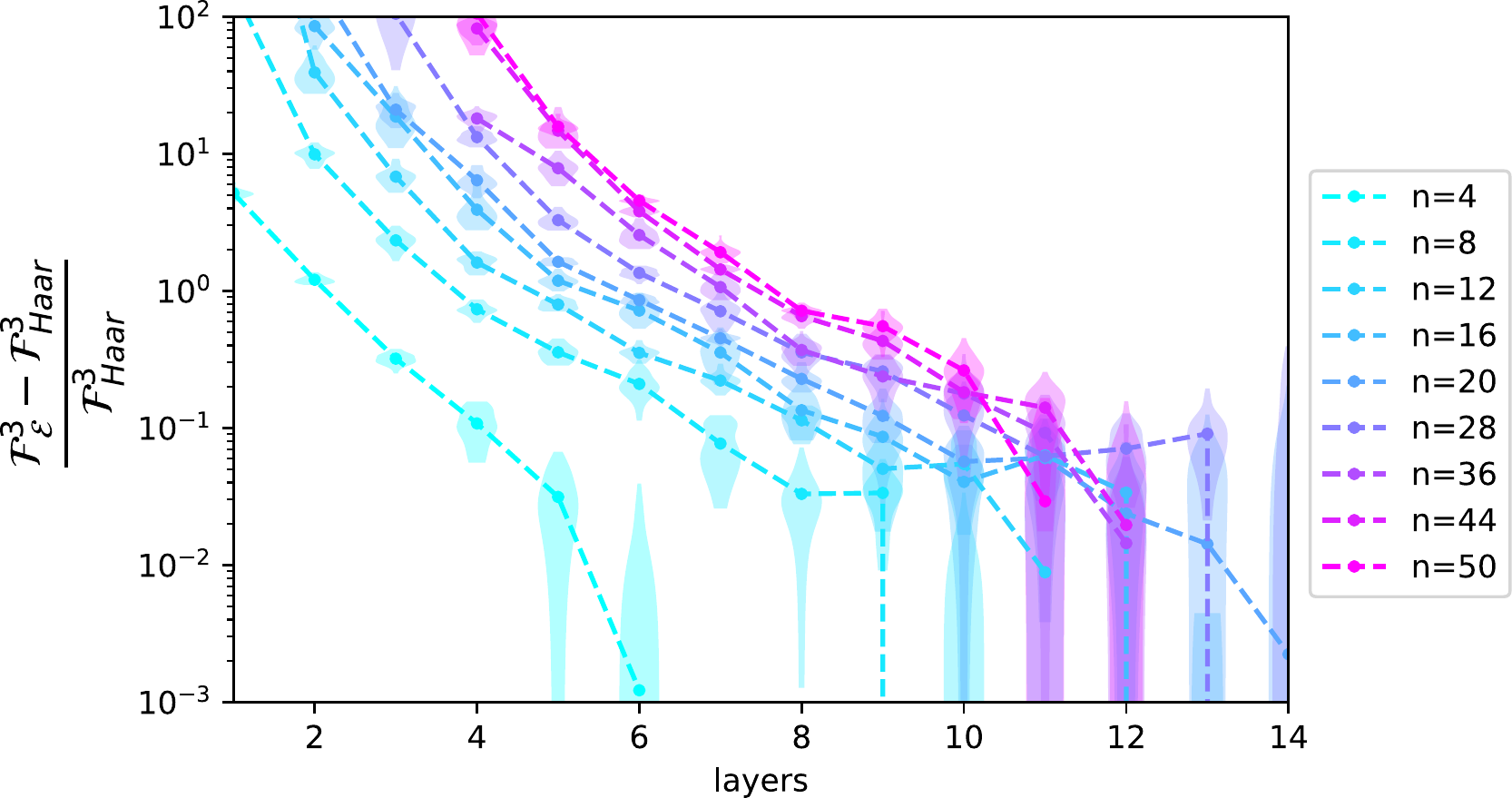}
   \end{center}
   \caption{Percentage deviation of the $k=3$ frame potential from the Haar value as a function of layers for the parallel random unitary ans{\"a}tz.}
   { \label{parallel bootstrap}
}
   \end{figure}

\begin{figure} [ht]
   \begin{center}
   \includegraphics[width=7cm]{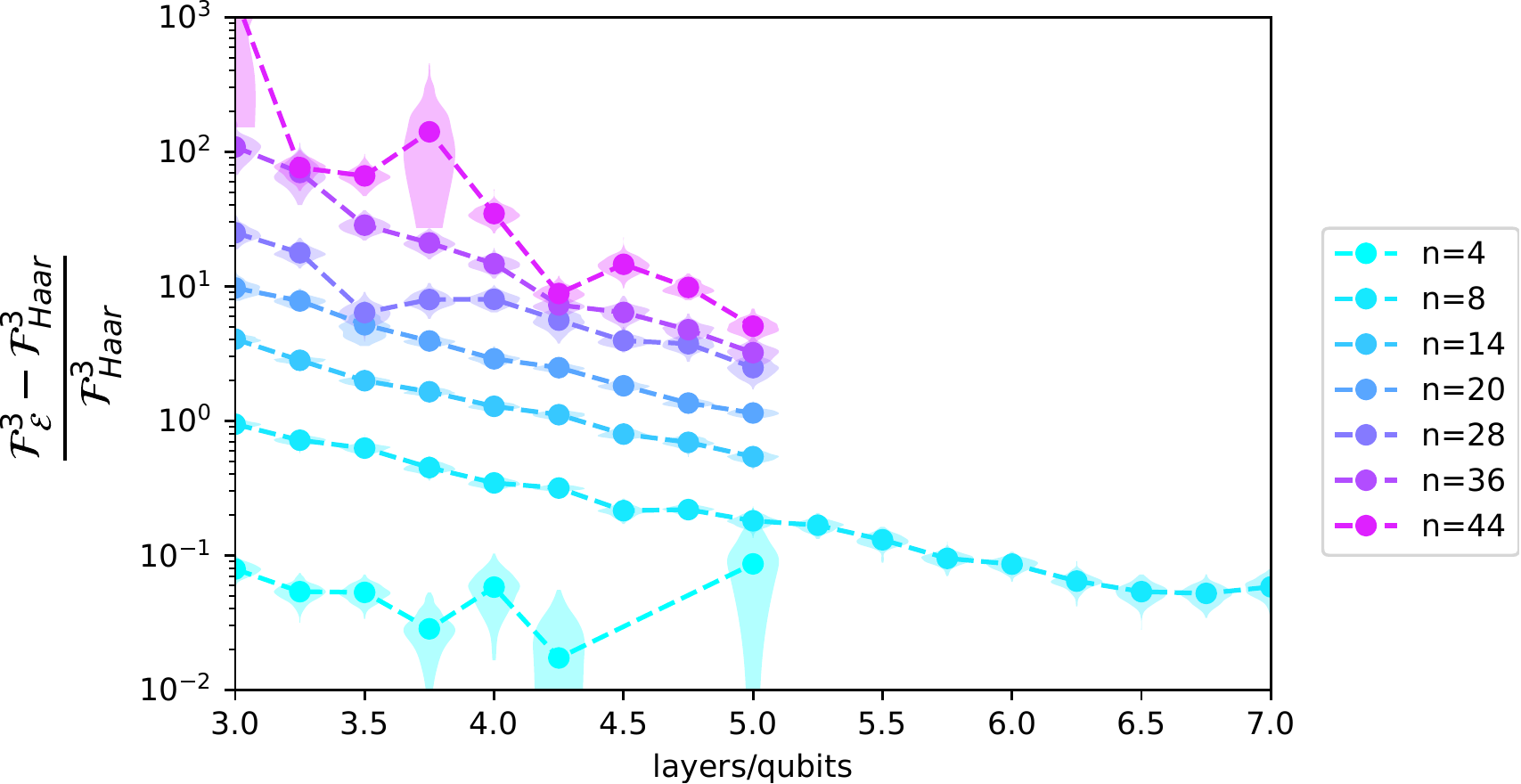}
   \end{center}
   \caption{Percentage deviation of the $k=3$ frame potential for the local random unitary ans{\"a}tz.}
   { \label{local bootstrap}
}
   \end{figure}

\begin{figure} [ht]
   \begin{center}
   \includegraphics[width=7cm]{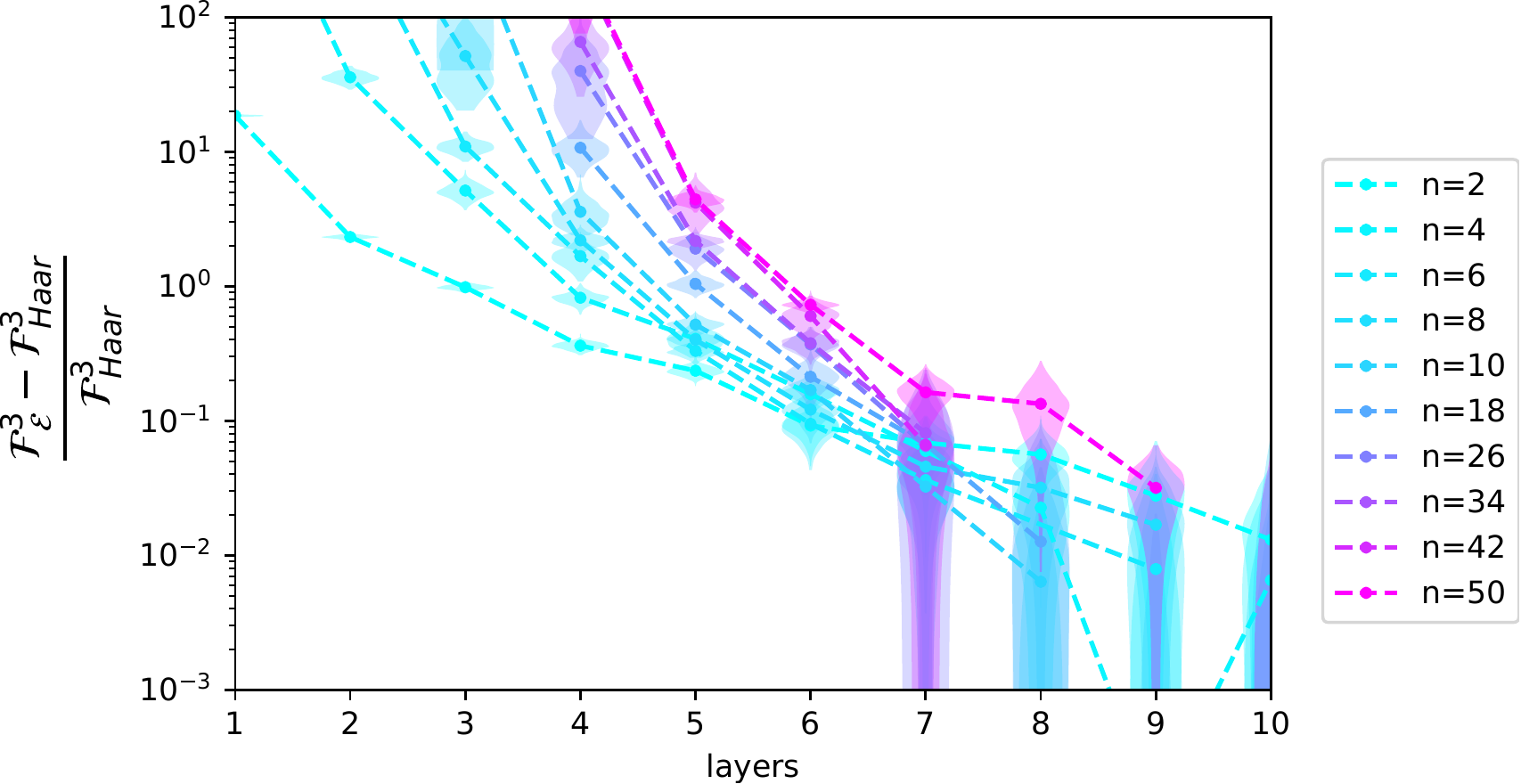}
   \end{center}
   \caption{Percentage deviation of the $k=3$ frame potential from the Haar value for the CNOT hardware efficient ans{\"a}tz.}
   { \label{HE bootstrap}
}
   \end{figure}

The main manuscript does not show uncertainties obtained from bootstrapping for the percentage deviation of the frame potential. This is because we want to prioritize illustrating traces of all qubit counts on the same plot, and bootstrapping errors can obfuscate the graphs. Supplementary Figure \ref{parallel bootstrap}, \ref{local bootstrap}, \ref{HE bootstrap} are violin plots that shows the bootstrap distributions of the frame potential percentage deviations as shadows around data points. To avoid cluttering the plots, we only show traces for half of the qubit counts.

\subsection{Validation of calculated frame potentials}

\begin{figure} [ht]
   \begin{center}
   \includegraphics[width=7cm]{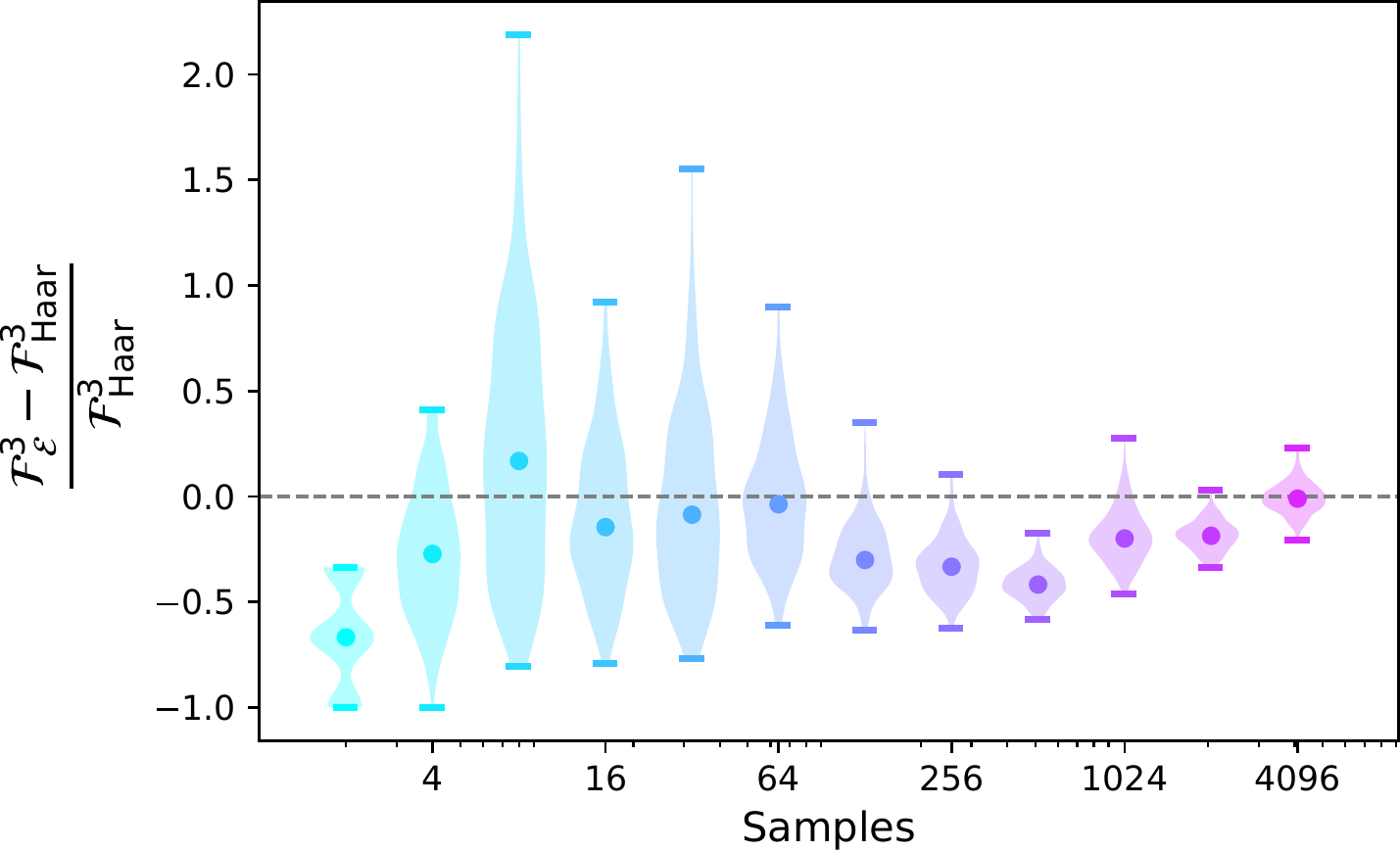}
   \end{center}
   \caption{Convergence process of the estimated frame potential as the number of samples increases. The plot shows the data for the parallel random unitaries with 50 qubits and 11 layers. Error analysis is similarly performed using bootstrapping. Solid points are medians of the bootstrap sample, and the vertical shadows represent the sample distribution where the width corresponds to the density. Horizontal bars show the extrema in the bootstrap samples.}
   { \label{convergence}
}
   \end{figure}

To show the convergence process, we take the parallel random unitary as an example, showing specifically the $n=50,l=11$ case, which is the largest number of qubits we simulated, and the largest number of layers simulated for this qubit count. We choose this as a weak data point since it is hard to simulate and obtain good statistics. Supplementary Figure \ref{convergence} shows the decrease in uncertainties as the number of samples increases. In this particular instance, the ans{\"a}tz is close to a 3-design.
   
\subsection{Validation of sampling Haar random two-qubit unitaries}

Parallel and local random unitaries require random two-qubit unitaries sampled from the Haar measure. The decomposition approach we use indeed satisfy this criterion. We observe for $n=2$ and $l=1$, which is a single random unitary gate, that the frame potential agrees with the Haar value $k!$ even at high $k$ values, verifying that it is sampling the Haar measure. Specifically, we show the convergence process for $k=6$ and $k=8$ in Supplementary Figure \ref{k=6}, \ref{k=8}.

\begin{figure} [ht]
   \begin{center}
   \includegraphics[width=7cm]{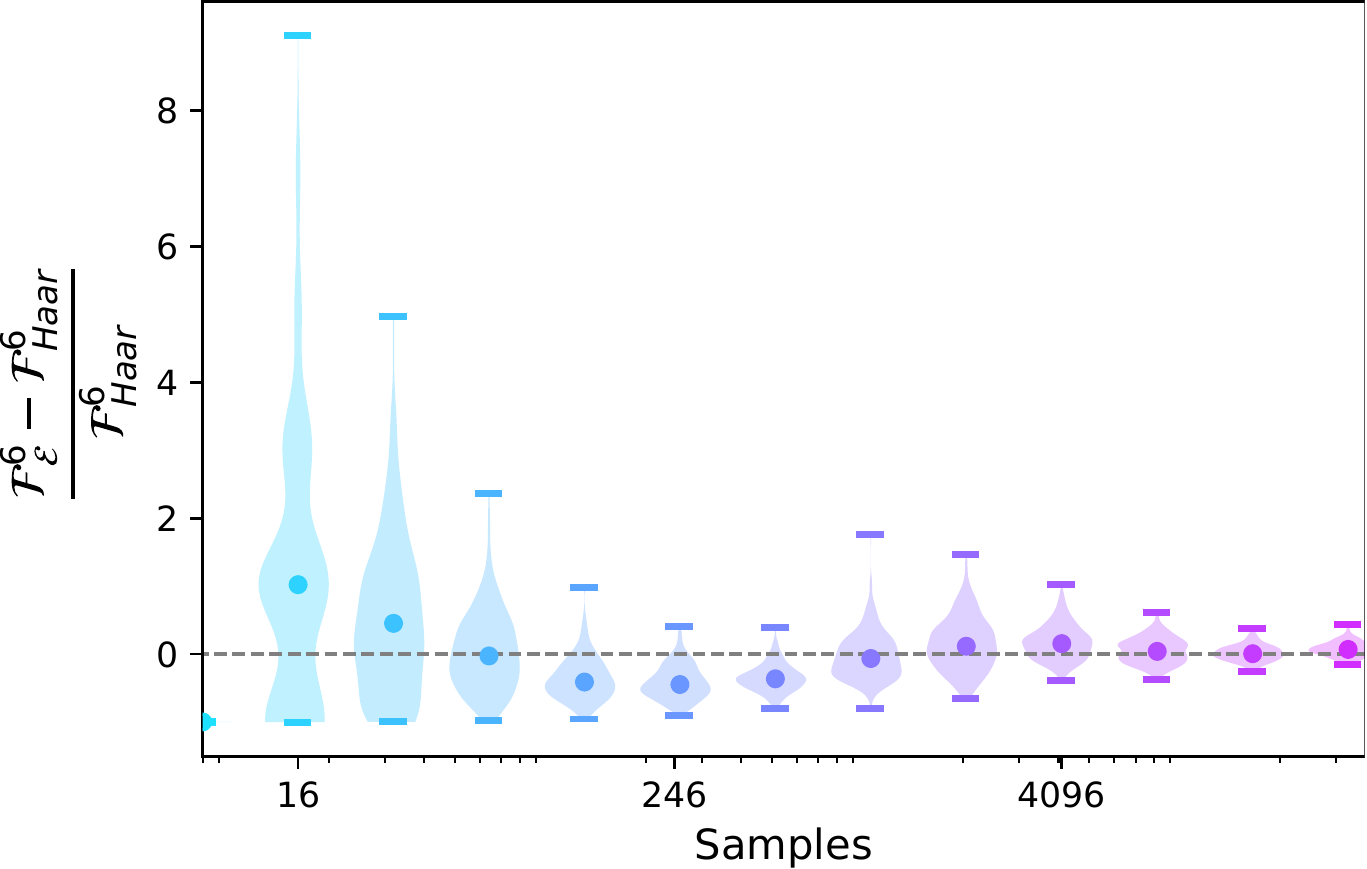}
   \end{center}
   \caption{Convergence of $\mathcal{F}$ for the two-qubit random gate to the Haar measure.} {\label{k=6}}
   \end{figure}
   
\begin{figure} [ht]
   \begin{center}
   \includegraphics[width=7cm]{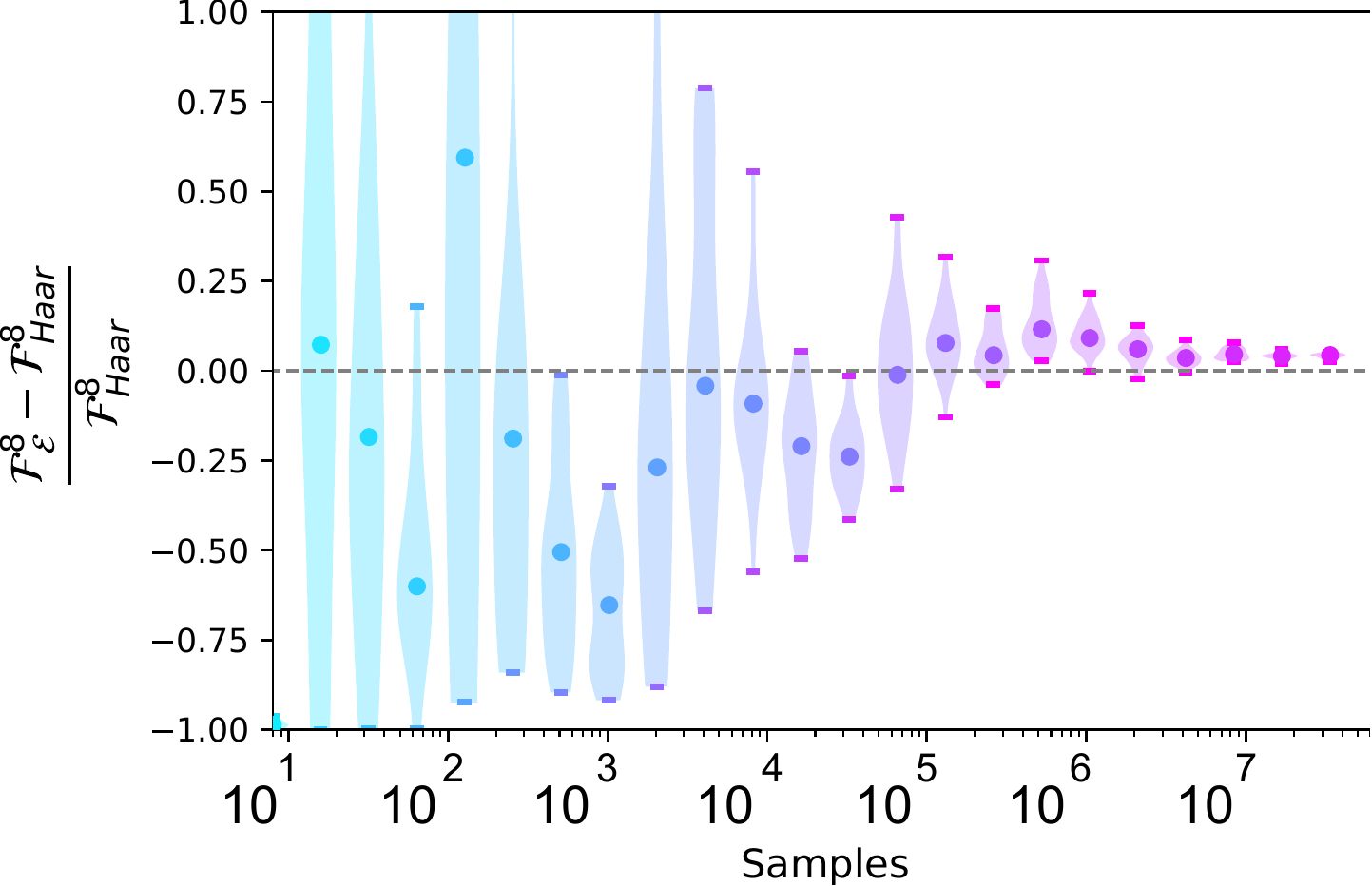}
   \end{center}
   \caption{Convergence for $k=8$.}  {\label{k=8}}
   \end{figure}

\section{Supplementary Discussion}

\subsection{Missing data points}

Increasing $k$ or $n$ leads to an increase in the frame potential percentage deviation from the Haar value, which means that the valid exponential regime starts later in $l$ for large $k$ and $n$. Further, due to the simulation difficulty of large $n$ circuits, we may not be able to simulate large $l$ and have to terminate the curves of $\frac{\mathcal{F}_\mathcal{E}^{(k)}-\mathcal{F}_{\text{Haar}}^{(k)}}{\mathcal{F}_{\text{Haar}}^{(k)}}\text{ v.s. }l$ early. This means that curves at large $k$ and $n$ are shorter in the exponential regime. Together with the larger uncertainties at large $k$ and $n$, there may not be enough data points for proper fitting, which is the reason why data points in the layer scaling plots (Figure 5, 7, and 9 of the main manuscript) are missing for large $k$ and $n$. One or more bootstrap samples for the missing data points fail to converge during the fit.

\subsection{Weak spurious correlation}

Since frame potentials of different $k$s are computed from the same set of $\text{Tr}\left(U^{\dagger}V\right)$, there is a correlation between different $k$'s. For example, the layer scaling for in Figure 5 and 9 of the main manuscript shows some weakly correlated fluctuations for both $k=2$ and $k=3$. However, if we split the data into three equal partitions and plot three curves, such fluctuations are no longer correlated, therefore do not represent genuine patterns.